\documentclass[]{autart}

\usepackage{amsmath}
\usepackage{bbold}
\usepackage{enumerate}
\usepackage{cite}
\usepackage{color,xcolor}
\usepackage{amssymb}
\usepackage{graphicx}
\usepackage{algorithm}
\usepackage{algpseudocode}
\usepackage{bbm}
\usepackage{graphics}
\usepackage{subfigure}
\usepackage{epsfig}
\usepackage{epstopdf}

\newtheorem{theorem}{Theorem}{}
{}
\newtheorem{lemma}{Lemma}{}
\newtheorem{assumption}{Assumption}
\newtheorem{corollary}{Corollary}{}
\newtheorem{definition}{Definition}{}
\newtheorem{remark}{Remark}{}

\begin{document}

\begin{frontmatter}

\title{Sensor Scheduling Design for Complex Networks under a Distributed State Estimation Framework\thanksref{mytitle}}
\thanks[mytitle]{The work by P. Duan, L. Huang and L. Shi is supported by a Hong Kong RGC General Research Fund 16206620. The work of L. He is supported by a Natural Science Foundation of China under Grant 61973163  and a Science Foundation of Jiangsu Province under Grant BK20191285. This work is accepted by {\it Automatica}. \emph{(Corresponding author: Lidong He)}}

\author[mymainaddress]{Peihu~Duan}\ead{eeduanpeihu@ust.hk},    
\author[mysecondaryaddress]{Lidong~He}\ead{lidonghe@njust.edu.cn},  
\author[mymainaddress]{Lingying Huang}\ead{lhuangaq@connect.ust.hk},               
\author[mythirdaddress]{Guanrong Chen}\ead{eegchen@cityu.edu.hk},               
\author[mymainaddress]{Ling Shi}\ead{eesling@ust.hk}  

\address[mymainaddress]{Department of Electronic and Computer Engineering, the Hong Kong University of Science and Technology, \\ Hong Kong SAR, China}

\address[mysecondaryaddress]{School of Automation, Nanjing University of Science and Technology, Nanjing 210094, China}

\address[mythirdaddress]{Department of Electrical Engineering, City University of Hong Kong, Hong Kong SAR, China}

\maketitle

\begin{abstract}
This paper investigates sensor scheduling for state estimation of complex networks over shared transmission channels. For a complex network of dynamical systems, referred to as nodes, a sensor network is adopted to measure and estimate the system states in a distributed way, where a sensor is used to measure a node. The estimates are transmitted from sensors to the associated nodes, in the presence of one-step time delay and  subject to packet loss. Due to limited transmission capability, only a portion of sensors are allowed to send information at each time step. The goal of this paper is to seek an optimal sensor scheduling policy minimizing the overall estimation errors. Under a distributed state estimation framework, this problem is reformulated as a Markov decision process, where the one-stage reward for each node is strongly coupled. The feasibility of the problem reformulation is ensured. In addition, an easy-to-check condition is established to guarantee the existence of an optimal deterministic and stationary policy. Moreover, it is found that the optimal policies have a threshold, which can be used to reduce the computational complexity in obtaining these policies. Finally, the effectiveness of the theoretical results is illustrated by several simulation examples.
\end{abstract}


\end{frontmatter}

\section{Introduction} \label{sec1}
Cyber-Physical Systems (CPSs) have been rapidly developed for industrial and military applications, supported by the well-developed technologies of sensing, communication and control \cite{baheti2011cyber,chen2012int,wang2019lq,lv2019adaptive}.

To guarantee satisfactory performances of CPSs, one critical issue is ensuring a reliable transmission of tremendous sensing data from  wireless sensors to remote estimators for state estimation \cite{li2015jamming,peihu2020}. As communication resources are always limited, it is important to synthesize a scheduler for sensors to achieve a better trade-off between the state estimation performance and the communication overhead.

The above problem of sensor scheduling for CPSs has attracted considerable attention from the research communities over the past two decades. According to   available information and  designing criteria, sensor scheduling can be divided into three typical classes, namely time-based scheduling \cite{mo2011sensor,ORIHUELA20142672}, event-based scheduling \cite{wu2012event,he2018event,han2015stochastic,YU2021109314}, and performance-based scheduling \cite{shi2011sensor,zhang7054460,Li7451220,LEONG201854}. First, as an offline strategy, the time-based scheduling policy is easy to design and implement. However, its performance is usually not as good as the other two online scheduling policies. In event-based scheduling, the sensor transmission is scheduled by events generated from a triggering condition, which is designed based on state or output variables. Generally, event-based policies can be classified into deterministic schedulers \cite{wu2012event,he2018event} and stochastic ones \cite{han2015stochastic,YU2021109314}. However, event-based policies are usually not optimal for a given performance index. To deal with this issue, a large number of works have been devoted to optimal scheduling, i.e., performance-based scheduling. As a first attempt, Shi \textit{et al.} \cite{shi2011sensor} designed an optimal sensor scheduler for state estimation with communication-energy constraint. Later, similar problems under other transmission constraints were inverstigated \cite{zhang7054460,Li7451220,LEONG201854}. In the present paper, optimal scheduling policies subject to a pre-desired performance index are proposed and developed.

It should be noted that the aforementioned works mainly studied a single process, i.e., only one sensor along with one remote estimator was considered. For optimal scheduling for multiple sensors, a solution to optimal scheduling two Gauss-Markov systems under bandwidth constraint was derived in \cite{shi2012scheduling}. Later, using a Markov decision process (MDP) approach, the result was extended to cases with multiple independent processes in \cite{han2017optimal,ren2018attack}. Recently, to make the MDP technique more computationally efficient, Whittle index-based techniques were applied in \cite{wang2019whittle,wu2020Optimal1}. Besides the MDP methods, a prioritizing scheduler based on the concept of Value of Information was designed in \cite{molin2019scheduling}. However, the multiple processes considered in these works must be independent (decoupled), which is hardly possible to directly apply the techniques in these works for coupled systems. As a remedy, joint estimation and scheduling for two interconnected systems was proposed in \cite{vasconcelos2019observation}. However, the observed system dynamics were omitted by assuming that they are zero-mean Gaussian random variables. Hence, the sensor scheduling for multiple interconnected systems has not been fully tackled, letting alone the cases with various transmission constraints.


In this paper, an optimal sensor scheduling scheme for state estimation of complex networks is studied, where the dynamics of nodes (subsystems) are strongly coupled. A sensor network is adopted to measure and estimate the node system states in a distributed way, where a sensor is associated with a node. In this setting, the transmitted data from sensors to nodes are correlated. Moreover, the case with packet losses and one-step time delay is considered. Under such transmission conditions, the problem is shown tractable, and some fundamental structures of optimal deterministic and stationary scheduling policies are found.

Compared to the literature, this paper possesses several advantageous features as follows:
\begin{enumerate}
  \item The sensor scheduling problem for complex networks is formulated as a standard MDP, based on a novel distributed state estimation framework with lower computational complexity compared to the augmented and linear matrix inequalities (LMIs)-based methods \cite{shen2010distributed,hu2016variance}.

  \item An easy-to-check condition is provided to ensure the existence of optimal deterministic and stationary policies for the  MDP problem, which only relies on the known system matrices.

  \item It is revealed that all optimal deterministic and stationary policies possess a threshold structure, which can be used to reduce the computation overhead in solving the MDP problem by avoiding global search of optimal policies. For the first time, this structure is found for complex networks.
\end{enumerate}

The remainder of this paper is organized as follows. In Section \ref{sec2}, some preliminaries and the problem statement are presented. In Section \ref{sec3}, the problem is reformulated as a standard MDP. Then, its feasibility  is analyzed and a structure of the optimal policy is revealed. In Section \ref{simulation}, several numerical examples are demonstrated to visualize the theoretical results. Finally, in Section \ref{Conclusion}, a conclusion is drawn.

\textit{Notations:}
For a  matrix $A \in \mathbb{R}^{n \times n}$, let  $A^T$, $A^{-1}$, $\text{Tr}(A)$ and $\rho(A)$ be the transpose, inverse, trace and spectral radius of $A$, respectively. For a positive-definite matrix $A \in \mathbb{R}^{n \times n}$, let $\lambda_{max}(A)$ represent the maximum eigenvalue of $A$.
For two matrices $A \in \mathbb{R}^{n \times n}$ and $B \in \mathbb{R}^{n \times n}$, let $A > (\ge) \ B$ mean that matrix $A-B$ is positive-definite (semidefinite). For a positive number $a \in  \mathbb{R}$, let  $\lceil a \rceil$ denote the least integer greater than or equal to $a$. For two matrices $S$ and $P$ of appropriate dimensions, let $SP(*)^T$ denote $SPS^T$. $\mathbb{Z}_{+}$ represents the set of all positive integers. Table \ref{table1} lists some frequently-used symbols.

\begin{table}[!htb]  \label{table1}
 \centering
Table 1. Some frequently-used symbols.
\begin{tabular}{c|l}
\hline
Symbol &  \qquad \qquad \qquad  Definition \\
\hline
$\breve{P}^{d}_{k,i}$ &  the minimized upper bound of $\mathbb{E} \{\overline{e}_{k,i} \overline{e}_{k,i}^T \}$   \\
\hline
$P^{d}_{k,i}$ &   the minimized upper bound of $\mathbb{E} \{e_{k,i} e_{k,i}^T \}$   \\
\hline
$\breve{P}_{i}$ & the steady state of $\breve{P}^{d}_{k,i}$, such as $\breve{P}_{1}$,  $\breve{P}_{j}$, $\breve{P}_h$   \\
\hline
$P_{i}$ & the steady state of $P^{d}_{k,i}$, such as $P_{1}$,  $P_{j}$, $P_h$ \\
\hline
$P_{k,i}$ & the index for node $i$ at step $k$, such as  $P_{k,h}$ \\
\hline
$P^{sum}_{k}$ & the sum of $P_{k,i}$, $i=1$, $\cdots$, $N$  \\
\hline
\end{tabular}
\end{table}

\section{Preliminaries and Problem Formulation}\label{sec2}

\subsection{Distributed State Estimation}\label{sec2.2}
In this subsection, state estimation for a complex network  consisting of $N$ subsystems (nodes) is considered, in which the node-system dynamics are described by
\begin{align} \label{system_model}
x_{k+1,i} & = A x_{k,i} + \mu \sum_{j=1}^{N} a_{ij} G  x_{k,j} + \omega_{k,i}, \\
\label{sensors} y_{k+1,i} & = C x_{k+1,i} + \nu_{k+1,i},  \ i=1, \ \ldots, \ N,
\end{align}
where $x_{k+1,i} \in \mathbb{R}^{n}$ and $y_{k+1,i} \in \mathbb{R}^{m} $ are the system state and the sensor measurement of node $i$ at step $k+1$, respectively. In this model, $A \in \mathbb{R}^{n \times n}$, $G \in \mathbb{R}^{n \times n} $ and $C \in \mathbb{R}^{m \times n}$ are system matrices and $\mu$ is the coupling parameter, $\omega_{k,i} \in \mathbb{R}^{n }$ and $\nu_{k+1,i} \in \mathbb{R}^{m}$ are Gaussian white noises with covariances $Q_i \ (> 0) \in \mathbb{R}^{n \times n} $ and $R_i \ (>0)  \in \mathbb{R}^{m \times m}$, respectively. The initial state $x_{0,i}$ is known as $\overline{x}_{0,i}$.  Besides, let $a_{ij} \in \{0, 1\}$ denote whether node $i$ can communicate with node $j$. Specifically, when $a_{ij}= 1$, node $i$ can receive information from node $j$, otherwise $a_{ij}= 0$. In particular, assume that there is no self-loop, i.e., $a_{ii}= 0$.  In this setting,  the neighbors set of node $i$ is denoted by $\mathcal{N}_i \triangleq \{  \ j \ | \ a_{ij} =1, \ j =1, \ \ldots, \ N \}$,  $d_i \triangleq | \mathcal{N}_i | $ and $d_{m} = \max \{d_i, \ i=1, \ \ldots, \ N\}$. Moreover, the adjacency matrix of the communication topology is denoted by $ \mathcal{A} \triangleq [a_{ij}]_{N \times N} \in R^{N \times N}$. In addition to interconnected systems, this model can also describe a large number of closed-loop multi-agent systems for consensus, formation and containment tasks \cite{gu2012consensus}.

This paper introduces a distributed state estimation algorithm for the complex network (\ref{system_model}) with sensors (\ref{sensors}). Following \cite{hu2016variance,li2018resilient,peihu2020}, the estimator is proposed as
\begin{align} \label{priorestimate}
\overline{x}_{k+1,i} & = A \tilde{x}_{k,i} + \mu \sum_{j=1}^{N} a_{ij} G  \tilde{x}_{k,j}, \\
\label{posterioriestimate} \tilde{x}_{k+1,i} & = \overline{x}_{k+1,i}  + K_{k+1,i} ( y_{k+1,i} - C \overline{x}_{k+1,i} ),
\end{align}
where $\overline{x}_{k+1,i} \in \mathbb{R}^{n} $ and $\tilde{x}_{k+1,i} \in \mathbb{R}^{n} $ are priori and posteriori estimates of ${x}_{k+1,i}$ by sensors, respectively, and $K_{k,i} \in \mathbb{R}^{n \times m} $ is the estimator gain to be designed. The priori and posteriori estimation errors for node $i$ at step $k$ are defined as $\overline{e}_{k,i} = \overline{x}_{k,i} - x_{k,i}$ and $e_{k,i} = \tilde{x}_{k,i} - x_{k,i}$, respectively.


%
%
%

\begin{definition} \label{definition1}
$(A, \ A_{0}, \ C)$ is detectable if there exists a matrix $K$ such that the following system
\begin{align}
x_{k+1} & = (I -  KC)  ( A  +   A_{0} w_k) x_{k}  \notag
\end{align}
is asymptotically mean-square stable, where $w_k \in \mathbb{R}$ is zero-mean Gaussian noise with covariance $1$.
\end{definition}

\begin{lemma} \label{lemma2}
For the complex network (\ref{system_model}) with sensors (\ref{sensors}), the estimator (\ref{posterioriestimate}) is unbiased. Moreover, the estimation error covariances are upper bounded by $P_{k+1}^d$ at step $k+1$, if estimator gains $K_{k+1,i}$, $i=1, \ \ldots, \ N$, are designed as
\begin{align} \label{Kdis}
K_{k+1,i}  = &  \breve{P}_{k+1,i}^d C^T (C \breve{P}_{k+1,i}^d C^T + R_i)^{-1},
\end{align}
where
\begin{align}
\label{pd} P_{k+1,i}^d  = & ( (\breve{P}_{k+1,i}^d)^{-1}  + C^T R_{i}^{-1} C )^{-1} ,     \\
\label{brevepd} \breve{P}_{k+1,i}^d = & \tilde{A} P_{k,i}^d  \tilde{A}^T   +   d_{i} \sum_{j \in \mathcal{N}_i} \tilde{G} P_{k,j}^d   \tilde{G}^T  + Q_i,
\end{align}
with $\tilde{A} = \sqrt{1 + \mu} A$ and $\tilde{G} = \sqrt{\mu + \mu^2} G$. Moreover, if $(\tilde{A},   \  d_{m}  \tilde{G}, \ C)$ is detectable, then $P_{k,i}^d$ is uniformly bounded and converges to a constant matrix exponentially.
\end{lemma}

Due to the space limitation, the proof of Lemma \ref{lemma2} is given in Appendix \ref{lemma2proof}.
The condition that $(\tilde{A}, \ d_{m}  \tilde{G}, \ C)$ is detectable can be replaced by a simpler one given in the following Lemma.

\begin{lemma} \label{lemma4}
When $(C, \ \tilde{A})$ is detectable, $P_{k,i}$ is uniformly bounded and convergent if
\begin{align} \label{condition}
\|G\|_2 \leq  \sqrt{\frac{z}{d_m^2(\mu+\mu^2)\lambda_{max}(\Theta)} }
\end{align}
 with
 \begin{align}
\Theta  = & (I_n - LC) ( \tilde{A} \Theta \tilde{A}^T + z I_n  + Q_{m} )(*)^T  + LR_mL^T,  \notag
\end{align}
where $L$ is any matrix ensuring that $(I_n - LC) \tilde{A}$ is Schur stable; $z$ is any positive scalar;  $R_{m}$ and $Q_{m}$ are chosen such that $R_{m} \ge   R_i  $ and  $Q_{m} \ge    Q_i, \ \forall i=1, \ \ldots, \ N$, respectively.
\end{lemma}

The proof of Lemma \ref{lemma4} is provided in Appendix \ref{lemma4proof}.

\begin{remark}
For distributed state estimation, two important cases are considered in the literature: 1) state estimation for a single system with multiple sensors where all sensors correspond to this same system \cite{olfati2007consensus}; 2) state estimation for a complex network with multiple sensors where each sensor corresponds to one node \cite{li2018resilient,shen2010distributed}. In this paper, the second case is investigated. Particularly, the ``distributed" concept here refers to that the estimator design only relies on local information and local interaction. On the one hand, the state estimate $\tilde{x}_{k,i}$ is shared among adjacent nodes for constructing the priori state estimator (\ref{priorestimate}). On the other hand, the recursive matrix $P_{k,i}^d$ in (\ref{brevepd}) is exchanged between adjacent nodes to design the estimator gain in a distributed manner. 
\end{remark}

\begin{remark}
Compared to centralized methods \cite{shen2010distributed,hu2016variance}, the designed algorithm (\ref{priorestimate})-(\ref{brevepd}) possesses advantages in computational efficiency, system robustness and scalability. First, according to \cite{arora2009computational}, the computational complexity of the algorithms in \cite{shen2010distributed,hu2016variance} is $O(N^3 n^3$ $ + N^3  m^3)$ while the one of the designed algorithm (\ref{priorestimate})-(\ref{brevepd}) is $O(d_i n^3 + m^3)$. Therefore, the designed algorithm has lower computational complexity. Second, the centralized methods heavily rely on a central authority to collect and fuse global information while the designed algorithm does not. By contrast, the designed algorithm can better suit to the situation where the   authority is broken or attacked. In this sense, it has stronger system robustness. Third, compared to global interaction in  centralized methods, local interaction between nodes in (\ref{priorestimate})-(\ref{brevepd}) makes it easier to introduce additional nodes into the network, which will not affect other nodes globally.  Hence, the algorithm (\ref{priorestimate})-(\ref{brevepd}) has a better scalability. Moreover, due to high  computational efficiency and local interaction, the algorithm (\ref{priorestimate})-(\ref{brevepd}) is of more energy-saving in computation and communication. For these advantages, the distributed framework (\ref{priorestimate})-(\ref{brevepd}) is adopted for the following sensor scheduling design.
\end{remark}

\begin{remark}
In this paper, two alternative conditions are provided to ensure the stability of the designed algorithm (\ref{priorestimate})-(\ref{brevepd}): 1) $(\tilde{A},   \  d_{m}  \tilde{G}, \ C)$ is detectable; 2) $(C, \ \tilde{A})$ is detectable with (\ref{condition}) being satisfied. Particularly, if there is no coupling between nodes, these conditions reduce to the one that $(C, \ A)$ is detectable. Compared to the stability conditions in \cite[Theorem 1]{li2018resilient}, \cite[Theorem 2]{hujun2020variance}, the second condition here is simpler since it only depends on system matrices and can be checked in advance while some parameters in the conditions in \cite{li2018resilient}, \cite{hujun2020variance} need to be specifically chosen or designed at each estimation step (e.g., $\bar{\xi}_1$ and $\bar{\xi}_2$ in  \cite{hujun2020variance}).
\end{remark}

In this paper, the upper bound $P_{k+1,i}^d$ is regarded as the estimation performance evaluation index for sensor/node $i$. Its steady state satisfies
\begin{align}
\label{steadypd} P_{i}  = & ( (\breve{P}_{i})^{-1}  + C^T R_{i}^{-1} C )^{-1}  ,     \\
\label{steadybrevepd} \breve{P}_{i} = & \tilde{A}  P_{i}  \tilde{A}^T   +   d_{i} \sum_{j \in \mathcal{N}_i}   \tilde{G}  P_{j}   \tilde{G}^T + Q_i.
\end{align}
Since $P_{k+1,i}^d$ converges exponentially, for simplicity, suppose that the distributed state estimation algorithm (\ref{priorestimate})-(\ref{brevepd}) has reached its steady state, i.e., $P_{k+1,i}^d = P_{i}$ and $ \tilde{x}_{k,i} = \hat{x}_{k,i}^s \triangleq \overline{x}_{k,i}  + K_{i} ( y_{k,i} - C \overline{x}_{k,i} )$, where  $K_{i}$ is the steady state of $K_{k,i}$ and $\overline{x}_{k,i} $ is given in (\ref{priorestimate}).

\subsection{Problem of Interest}\label{sec2.3}
\begin{figure}[!htb]
\center
\subfigure{{\includegraphics[scale=0.5]{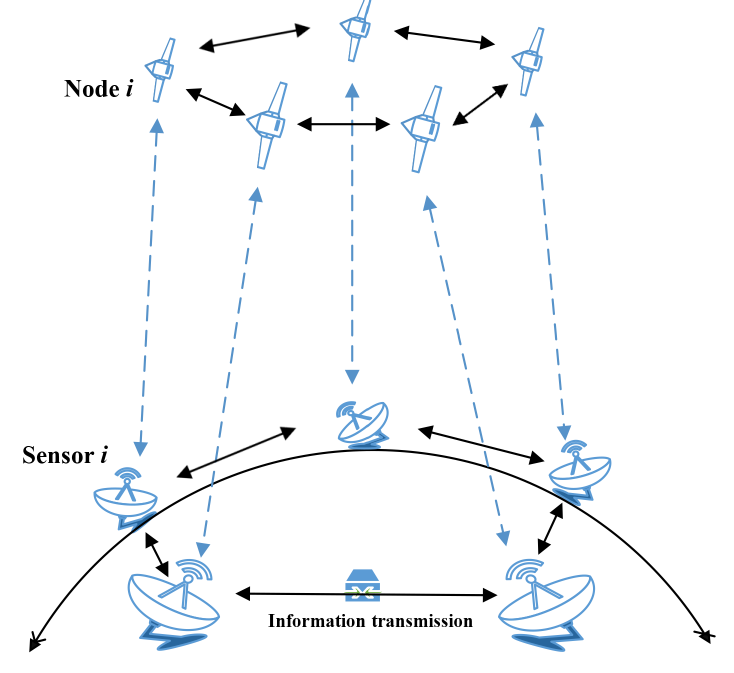}}}
\caption{An example of remote state estimation over a communication network.}  \label{f:plant}
\end{figure}

\begin{figure}[!htb]
\center
\subfigure{{\includegraphics[scale=0.3]{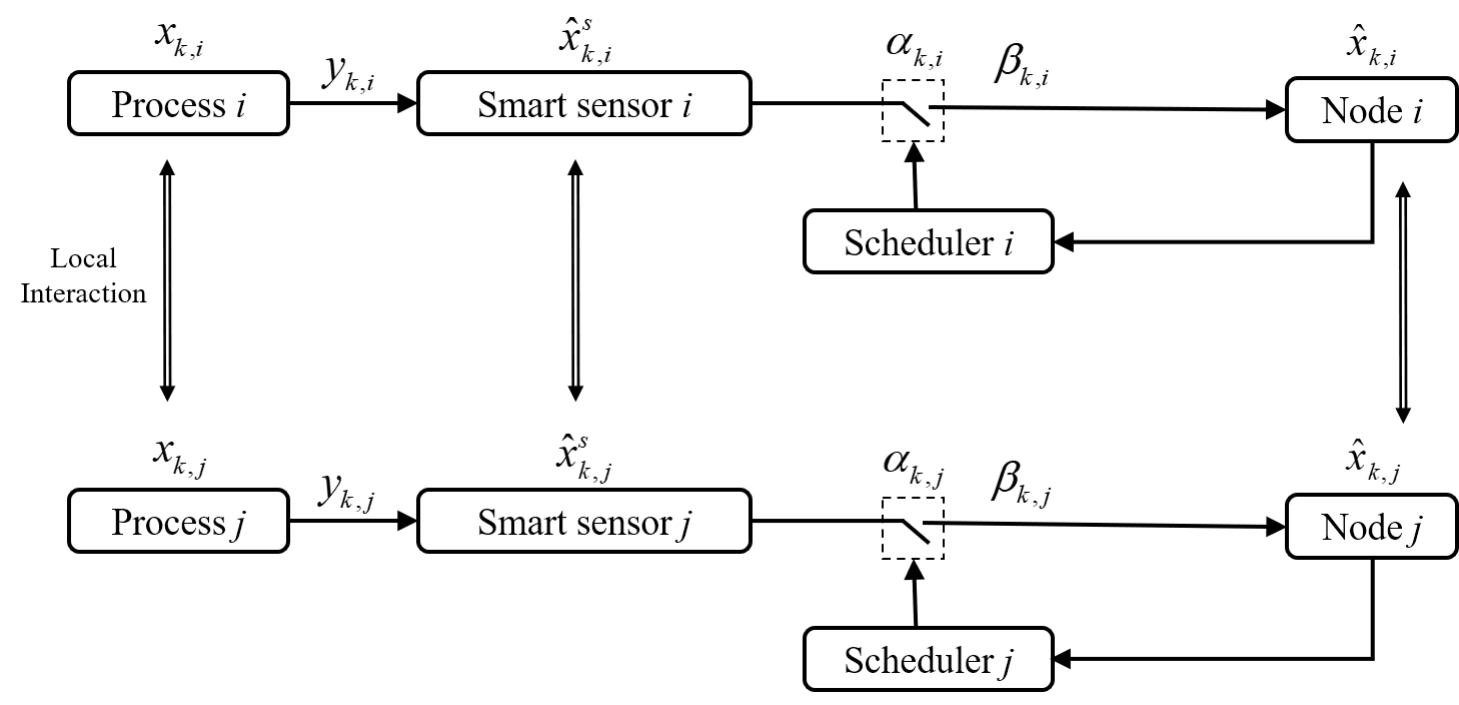}}}
\caption{The architecture of information transmission.}  \label{f:architecture}
\end{figure}

This paper considers remote state estimation for a complex network, where each node  is equipped with a smart sensor that has some computational capability. An illustrative example for such formulation is presented in Fig.~\ref{f:plant}, where multiple ground-based radars provide state estimates for unmanned aerial vehicles (UAVs) to complete a formation mission. Specifically, a radar is used to observe and estimate the state of a corresponding UAV, and the state estimate is transmitted to the corresponding UAV. The architecture of information transmission is shown in Fig.~\ref{f:architecture}, which consists of three communication counterparts: a node network, a sensor network and a sensor-node transmission network.

\textbf{Node network:} A network of coupled nodes with dynamics described by (\ref{system_model}) is assigned to perform cooperative tasks, where each node needs to obtain its current state information. In this network, every node can communicate with its neighbors.

\textbf{Sensor network:} A network of sensors with dynamics described by (\ref{sensors}) is  placed to measure the states of the nodes. To achieve an overall trade-off between the estimation performance and the communication overhead, the distributed state estimation framework (\ref{priorestimate})-(\ref{posterioriestimate}) is adopted for each sensor to estimate the state of its corresponding node.

\textbf{Sensor-node transmission network:} After computing the estimation, sensors need to send the estimates to nodes timely. This transmission process possesses the following features:

1) \textbf{Limited transmission capability:} Considering that the wireless transmission systems from sensors to nodes are homogeneous, all sensor-node transmissions share a common or similar channel. To avoid data interference, up to $M$ out of the $N$ sensors can transmit their state estimates, simultaneously.

2) \textbf{Uncertain transmission process:} Let $\alpha_{k,i} \in \{0, \ 1\}$ denote whether node $i$ requires sensor $i$ to transmit the state estimate at step $k$. When the transmission is required, $\alpha_{k,i} = 1$; otherwise $\alpha_{k,i} = 0$. There exist uncertainties in the data transmission processes, which are assumed to be Bernoulli packet-loss processes. Let $\beta_{k,i} =0$ represent that the packet is lost; $\beta_{k,i} =1$ otherwise. Moreover, denote the successful transmission rate from sensor $i$ to the associated node by $\lambda_i$, i.e., $\mathbb{E} \{ \beta_{k,i} \} = \lambda_i$. In addition, let $\text{cov} ( \beta_{k,i}, \ \beta_{k,j} ) = 0$, $ \forall i \neq j$.

3) \textbf{One-step transmission delay:} Since the distance between sensors and nodes might be rather long, for example between ground-based radars and satellites, suppose that one-step time delay occurs in the data transmission processes.

Let $\gamma_{k,i} \in \{0, \ 1\} $ represent the arrival of the packet $\hat{x}_{k,i}^s$ at step $k+1$. If the packet is received, $\gamma_{k,i}=1$; otherwise $\gamma_{k,i}=0$.  For node $i$, denote the time duration from the last step, when a packet was successfully received, to the current step $k$  by $\tau_{k,i} $, which is defined as follows:
\begin{align} \label{tau}
\tau_{k,i} = k - \text{max}_{l} \{l|\gamma_{l,i}=1, l < k   \}.
\end{align}

To maximize the transmission capacity of the network, when sensor $i$ is allowed to transmit information at step $k$, it transmits the packet $\{\hat{x}_{l,i}^s\}$, $l=k-\tau_{k,i}+1,\ldots,k$, to node $i$. Then, due to the existence of one-step time delay in the transmission processes, nodes need to compute the state estimates at step $k+1$ with the information at step $k$. In this paper, the prior-estimator (\ref{priorestimate}) is adopted, i.e.,
\begin{align}
& \hat{x}_{k+1,i} =  A X_{k,i}  + \mu \sum_{j \in \mathcal{N}_i} a_{ij} G  X_{k,j},   \notag
\end{align}
with
  \begin{align}
    X_{l,h} = &
  \begin{cases}
   \hat{x}_{l,h}^s,    & \text{if $ \gamma_{k,h}= 1$  },  \\
   \hat{x}_{l,h},    & \text{if $ \gamma_{k,h} =  0$ }, \qquad  \\
  \end{cases} \notag
  \end{align}
where $h = i$ or $j$, and $l=k-\tau_{k,i}+1$, $\ldots$, $k$. On the basis of Lemma \ref{lemma2}, the corresponding performance evaluation index is chosen as
\begin{align} \label{Piteration}
& P_{k+1,i} =   h_i(Y_{k,i}, Y_{k,j}, j \in \mathcal{N}_i),
\end{align}
with
\begin{align}
&h_i(Y_{k,i}, Y_{k,j}, j \in \mathcal{N}_i) =
    \tilde{A} Y_{k,i}  \tilde{A}^T   +   d_{i} \sum_{j \in \mathcal{N}_i}  \tilde{G} Y_{k,j}   \tilde{G}^T + Q_i ,    \notag
 \end{align}
and
 \begin{align}
   Y_{l,h} = &
  \begin{cases}
   P_{h},    & \text{if $ \gamma_{k,h}= 1, $ }  \\
   P_{l,h},     & \text{if $ \gamma_{k,h}= 0, $ }
  \end{cases} \notag
  \end{align}
where $h = i$ or $j$, $ l=k-\tau_{k,i}+1, \ \ldots, \ k-1, \ k $, and $P_{h}$ and $P_{l,h}$ are defined the same with $P_{i}$ and $P_{k,i}$, respectively.
\begin{remark}
Compared to the strategy of transmitting raw measurements, the strategy of performing distributed estimation in advance by smart sensors and then transmitting state estimates to nodes is more preferable, since the loss of transmitted measurements in the former strategy will have an unerasable effect on the estimation performance at future steps while the loss of transmitted state estimates in the latter strategy will not be so, as soon as the future state estimate is received. Actually, the latter strategy has been widely adopted in existing works on sensor scheduling for a single system or multiple independent systems \cite{shi2012scheduling,han2017optimal,wu2020Optimal1}. Here, this idea is extended to complex networks.
\end{remark}

Now, the focus is on the overall system performance under a scheduling policy $\Xi = \{\alpha_{k,i}\}$, $i=1, \ \ldots, \ N, \ k=1, \ \ldots,$ which is defined as an expected average cost
\begin{align} \label{j_theta}
& J (\Xi) \triangleq  \lim_{T \rightarrow  \infty} \text{sup} \frac{1}{T} \mathbb{E} \bigg [  \sum_{k=1}^{T} \sum_{i=1}^{N} \big ( \text{Tr}(P_{k,i}) +  \kappa \alpha_{k,i} \big ) \bigg].
\end{align}
The cost function $J(\Xi)$ defined in (\ref{j_theta}) consists of estimation performance indices $\text{Tr}(P_{k,i})$ and scheduling costs $\kappa \alpha_{k,i}$. On the one hand, in the standard Kalman filtering for a single system, the minimized $Tr(P_{k} \triangleq \mathbb{E}\{e_{k} e_{k}^T \} )$ guarantees the optimal estimation performance (e.g., the minimized mean-square estimation error), where $e_k$ is the estimation error. In this paper,  this idea is extended to complex networks, adopting $Tr(P_{k,i})$ as the estimation performance index for node $i$ at step $k$. On the other hand, since a scheduling operation usually costs a scheduling resource, the positive scalar $\kappa$ is introduced to measure the cost of a scheduling operation. To simultaneously evaluate the estimation performances and the scheduling costs of all nodes at all steps,  $J(\Xi)$ is designed as (\ref{j_theta}). It is expected  to find an optimal scheduling policy that minimizes this cost, as formulated below:

$\textbf{Problem 1}$
\begin{align}
 & \qquad \min_{ \Xi } J (\Xi), \notag \\
  s.t. \  & \sum_{i=1}^{N} \alpha_{k,i}=M, \ \forall k \ge 1. \notag
\end{align}

\begin{remark}
Compared to relevant problems investigated in \cite{han2017optimal,wu2020Optimal1}, $\textbf{Problem 1}$ possesses two key coupled features simultaneously. First, the performance index $P_{k,i}$ depends heavily on the scheduling indices of neighbors at the previous step, i.e., $P_{k-1,j}$, $j \in \mathcal{N}_i$. Specifically, for node $i$, the neighbors' scheduling signals $\alpha_{k-1,j}$ directly affect the value of $P_{k,i}$, which further has an influence on  $\alpha_{k,i}$. Second, the total scheduling resource at each step is limited, i.e., $\sum_{i=1}^{N} \alpha_{k,i}=M < N$, which also causes strong system coupling. The above formulation establishes a basic framework for sensor scheduling of complex networks, which can be extended to networks with non-identical matrices. For example, it is also possible to find an optimal scheduling policy for a network with heterogeneous nodes \cite{liu2021moving} by solving $\textbf{Problem 1}$.
\end{remark}

\section{Main results} \label{sec3}
In this section,  $\textbf{Problem 1}$ is reformulated as a standard MDP problem. Then, a sufficient condition is established to ensure the existence of a deterministic stationary optimal policy for this MDP problem. To that end, a structural property of optimal policies is revealed, which can be used to greatly reduce the computational cost for solving the MDP problem.

\subsection{MDP Formulation} \label{sec3.1}

First, $\textbf{Problem 1}$ is formulated as a standard MDP problem, described by a tuple $\{\mathcal{S}, \ \mathcal{A}, \ \mathcal{P}(\cdot|\cdot,\cdot), \ c(\cdot,\cdot)\}$:
\begin{enumerate}
  \item The state space $\mathcal{S} = \mathbb{Z}^N_{+}$ is the set of all possible states $s_{k} = \{\tau_{k,1}, \ \ldots, \ \tau_{k,N}\} $, where $\tau_{k,i}$, $i=1$, $\ldots$, $N$, is the time duration defined in (\ref{tau}).
      \\
  \item The action space $\mathcal{A} = \{0, \ 1\}^N$ is the set of all possible scheduling actions $a_{k} =  \{\alpha_{k,1}, \ \ldots, \ \alpha_{k,N}\} $, where $\alpha_{k,i}$ is the scheduling signal given in  the uncertain transmission process part  in Section \ref{sec2.3}.
      \\
  \item The transition kernel $\mathcal{P}(\cdot|\cdot,\cdot)$ is the transition probability from state $s_{k}$ to state  $s_{k+1}$ under action $a_{k}$, i.e.,
  \begin{align} \label{probability}
 \mathcal{P}(s_{k+1}|s_{k},a_{k}) = \prod_{i=1}^{N} \text{Pr}(\tau_{k+1,i}|\tau_{k,i},\alpha_{k,i}),
  \end{align}
  where
  \begin{align}
  & \text{Pr}(\tau_{k+1,i}|\tau_{k,i},\alpha_{k,i}) \notag \\
  = & \begin{cases}
    \lambda_i,    & \text{if $ \tau_{k+1,i} = 1$, $\alpha_{k,i} = 1$}, \\
    1-\lambda_i,     & \text{if $ \tau_{k+1,i} = \tau_{k,i}+1$, $\alpha_{k,i} = 1$}, \\
     1,    & \text{if $ \tau_{k+1,i} = \tau_{k,i}+1$, $\alpha_{k,i} = 0$}, \\
     0,   & \text{otherwise}.
  \end{cases} \notag
  \end{align}
  for all $i=1$, $\ldots$, $N$. Just for concise presentation, assume that $\lambda_i$ is identical for all nodes, i.e.,  $\lambda_i = \lambda$, $i=1$, $\ldots$, $N$.

  \item The one-stage reward  $ c(\cdot,\cdot)$ is defined as
    \begin{align}
   c(s_{0:k},a_{k})  =   & \sum_{i=1}^{N}  c_i(s_{0:k},\alpha_{k,i}), \notag
  \end{align}
where $ c_i(s_{0:k}, \alpha_{k,i})  =   \text{Tr} ( h_i(Y_{k-1,i}, Y_{k-1,j}, j \in \mathcal{N}_i)) $ $  +  \kappa \alpha_{k,i}$, $s_{0:k} \triangleq \{s_h\}_{h=0}^{k}=\{s_0,s_1,\ldots,s_k\}$, and $Y_{k-1,i}$ and $Y_{k-1,j}$ are defined in (\ref{Piteration}).

\end{enumerate}

\begin{theorem} \label{proposition1}
$ c(s_{0:k},a_{k})$  is independent of $s_{0:k-1}$.
\end{theorem}

The proof of Theorem \ref{proposition1} is provided in Appendix \ref{proposition1proof}. Theorem \ref{proposition1} indicates that a given pair of ($s_{k}$,  $\alpha_{k,i}$) generates a deterministic one-stage reward, which renders the MDP problem reformulation feasible. Since $ c(s_{0:k},a_{k})$ is determined by $s_{k}$ and $a_{k}$, one can re-write it as $ c(s_{k},a_{k})$.

Then, a feasible policy $\pi$ that corresponds to the scheduling policy is a sequence of functions $\{\pi_1, \ldots, \pi_k, \ldots\}$, which maps the history $(s_{0:k}, a_{0:k-1})$ to the action space, i.e., $a_k=\pi_{k}(s_{0:k},a_{0:k-1})=\{\pi_{k,i}(s_{0:k},a_{0:k-1})\}_{i=1}^{N}$. For all   feasible policies, the expected average cost is defined as
\begin{align}
& J (s_0,\pi) \triangleq  \lim_{T \rightarrow  \infty} \text{sup} \frac{1}{T} \mathbb{E}_{s_0}^{\pi} \bigg [  \sum_{k=1}^{T}  c(s_{k},\pi_{k}) \bigg]. \notag
\end{align}
Hence, $\textbf{Problem 1}$ can be reformulated as follows:

$\textbf{Problem 2}$
\begin{align}
 &   \min_{ \pi } J (s_0,\pi). \notag
\end{align}

\begin{remark}
It is worth mentioning that the relation between $c(s_{k}, \alpha_{k})$ and $s_{k} $  is complex, which cannot be expressed like the one in \cite{ren2018attack} (the sum of $N$ independent terms where each is only concerning a corresponding $\tau_{k,i}$, $i=1,\ldots, N$).  This factor results in great technical challenges for solving the above MDP problem. Besides, when the scale of the system is small, one can resort to the value iteration method to find the optimal policy. However, when the number of nodes is large, this traditional method fails due to the exponentially-increasing computational complexity.
\end{remark}

\subsection{Feasibility Conditions} \label{sec3.2}

In this subsection, a sufficient condition is established to ensure the existence of a deterministic and stationary optimal solution (policy) to $\textbf{Problem 2}$. Here, a policy is stationary if it is independent of time $k$, and a policy is deterministic if it is fully determined by the current MDP state.

\begin{assumption} \label{feasibilitycondition}
$(1-\lambda) \rho^{2r} < 1$, where $r=\lceil\frac{N}{M}\rceil$ and $\rho=\rho(U)$ with $U$ being any matrix that satisfies $U^T U = \tilde{A}^T \tilde{A} $ $ +   d_m^2  \tilde{G}^T  \tilde{G} $.
\end{assumption}

\begin{theorem} \label{thm1}
If Assumption \ref{feasibilitycondition} holds, there exists a deterministic stationary policy $\pi^{*}(\cdot)$, a constant $J^{*}$ and a function $V(\cdot)$ such that the following Bellman optimality equation holds for all $s \in \mathcal{S}$:
\begin{align} \label{bellmanopt}
 & J^{*} + V(s) =  \min_{ a \in  \mathcal{A} } \Big [ c(s,a) + \sum_{s^{+} \in \mathcal{S} } V (s^{+}) \mathcal{P}(s^{+}|s,a) \Big ] \notag \\
  = \ &  c(s,\pi^{*}(s)) + \sum_{s^{+} \in \mathcal{S} } V(s^{+}) \mathcal{P}(s^{+}|s,\pi^{*}(s)).
\end{align}
 Moreover, $ J^{*} =  J  (s_0,\pi^{*}) = \min_{ \pi } J (s_0,\pi) . $
\end{theorem}

The proof of Theorem \ref{thm1} is provided in Appendix \ref{thm1proof}. To prove Theorem \ref{thm1}, the existence of a deterministic stationary policy is formulated as the uniform boundedness of the average values of $\sum_{i=1}^{N}\text{Tr}(Y_{k,i})$ over the whole time horizon. Then, the latter is guaranteed by a modified round-robin scheduling policy using the condition in Assumption \ref{feasibilitycondition}.

\begin{remark}
Compared to the sufficient condition based on a recursive algorithm in \cite{wu2020Optimal1}, the condition in Assumption \ref{feasibilitycondition} is easier to satisfy. In particular, when $N=M$ and no coupling exists in the system dynamics, this condition collapses to $(1-\lambda) \rho^2(A) < 1$, which is a sufficient and necessary condition for independent processes in \cite{wu2020Optimal1}.
\end{remark}

\subsection{Structure of Optimal Policies} \label{sec3.3}
In this subsection, to greatly reduce computation cost in solving $\textbf{Problem 2}$,  a structural property of optimal policies is presented.

Since we look for a deterministic and stationary policy, the time index $k$ for $s_{k}$ and $\tau_{k,i}$ will be omitted for notational simplicity. Then, let $s_{i}$ denote $\tau_{i} $, and $s_{i}^{-}$ denote $ \{\tau_{1}, \ \ldots, \tau_{i-1}, \ \tau_{i+1},  \ \ldots, \ \tau_{N}\}  $, respectively. For any two states $s$ and $s'$, let $s \leq_i s'$ represent  $s_{i} \leq s_{i}'$ and  $s_{i}^{-} = s_{i}'^{-}$. Same definitions are made for $a =  \{\alpha_{i}\}_{i=1}^{N} $ and a series of square matrices $Z_{l} = \{Z_{l,h}   \in \mathbb{R}^{n \times n} \}_{h=1}^{N}$.

Before moving on, a lemma concerning the iteration of the one-step reward is presented as a preliminary result.

\begin{lemma} \label{lemmathm2}
For two series of matrices, $Z_{l}^{0}= \{Z_{l,h}^{0}  \in \mathbb{R}^{n \times n} \}_{h=1}^{N}$ and $Z_{l}^{1}= \{Z_{l,h}^1  \in \mathbb{R}^{n \times n} \}_{h=1}^{N}$, $l = k-\tau_i, \ \ldots, \ k$, where
 \begin{align}
   Z_{l,h}^{0} = &
  \begin{cases}
   P_{h},    & \text{if $ l \leq k - \tau_{h} $, }  \\
    h_h(Z_{l-1,h}^{0}, Z_{l-1,j}^{0}, j \in \mathcal{N}_h),  & \text{if $ l > k - \tau_{h},$ }
  \end{cases} \notag
  \end{align}
and
 \begin{align}
   Z_{l,h}^{1} = &
  \begin{cases}
   P_{h},    & \text{if $ l \leq k - \tau_{h} $, }  \\
    h_h(Z_{l-1,h}^{1}, Z_{l-1,j}^{1}, j \in \mathcal{N}_h),  & \text{if $ l > k - \tau_{h},$ }
  \end{cases} \notag
  \end{align}
if $Z_{k - \tau_{i}}^{0} \leq_i Z_{k - \tau_{i}}^{1}$, then $Z_{k,h}^{0} \leq Z_{k,h}^{1}$, $\forall h = 1, \ldots, N$.
 \end{lemma}

The proof of Lemma \ref{lemmathm2} is provided in Appendix \ref{lemmathm2proof}. Lemma \ref{lemmathm2} guarantees that the monotonicity of $Y_{k,i}$ with respect to $Y_{l,i}$, $\forall l < k$, is preserved by the operation of (\ref{Piteration}). It will serve as a preliminary result to prove the monotonicity of the one-stage reward $ c(s,a)$ concerning $s$, which will facilitate to reveal the structure of optimal policies.

\begin{theorem} \label{thm2}
If Assumption \ref{feasibilitycondition} holds, the optimal policy $\pi^{*}$ has a threshold structure, i.e., there exists a measurable function $\delta_i(\cdot)$ for node $i$, $ i =1, \ldots, N$, such that
\begin{align}
 & \pi^{*}_i (s) =    \begin{cases}
   1,    & \text{if $ s_{i} \ge \delta_i(s_{i}^{-}) $ }, \\
   0,    & \text{if $ s_{i} < \delta_i(s_{i}^{-}) $ }. \\
  \end{cases}  \notag
\end{align}
\end{theorem}

The proof of Theorem \ref{thm2} is provided in Appendix \ref{thm2proof}. The structure in Theorem \ref{thm2} indicates that, for an optimal policy, if the transmission is required for sensor $i$ when the state is $s$, the transmission must be required for sensor $i$ when the state is $s'$, $\forall s \leq_i s'$.  The policy with this threshold structure can facilitate online implementation of sensor scheduling by reducing the storage space.  Specifically, every node decides whether the corresponding sensor should transmit its information based only on the current state.

By leveraging this structure, a modified relative value iteration algorithm is proposed to reduce the computation cost in solving {\bf Problem 2}. Before moving on, define a series of functions for $s  \in \mathcal{S}$ as follows:
\begin{align}
 \label{v_function}     V_{k+1}(s) & =  \min_{ a \in  \mathcal{A} } \Big [ c(s,a) + \sum_{s^{+} \in \mathcal{S} } V_k (s^{+}) \mathcal{P}(s^{+}|s,a) \Big ],
\end{align}
where $k=0, \ 1, \ \ldots$. In addition, denote
\begin{align}
\label{optimalpi} \pi_{k+1}^{*}(s) & =  \arg \min_{ a \in  \mathcal{A} } \Big [ c(s,a) + \sum_{s^{+} \in \mathcal{S} } V_k (s^{+}) \mathcal{P}(s^{+}|s,a) \Big ].
\end{align}
 Then, a novel method for solving {\bf Problem 2} is designed in Algorithm \ref{algorithm1}.

 In the traditional brute-force methods such as relative value iteration (Algorithm \ref{algorithm1} without steps $5-7$, $9$) \cite{zhou2017optimal},  $V_{k+1}(s)$ and $\pi_{k+1}^{*}  (s)$ have to be computed by (\ref{v_function})-(\ref{optimalpi}) for all $s \in \mathcal{S} $, i.e., step $8$ in Algorithm \ref{algorithm1} has to be performed for all $s \in \mathcal{S} $. In this paper, by leveraging the threshold structure revealed in Theorem \ref{thm2}, steps $5-7$, $9$ are added into Algorithm \ref{algorithm1} such that step $8$ is not needed to run for states ``$s$'' satisfying $ s' \leq_i s$, where $s'$ is a state with $\pi_{k+1}^{*}  (s') = \bar a \in \mathcal{A} $ and $\bar a_i = 1$. It is worth mentioning that step $8$ is a computationally expensive step. By reducing the time of performing step $8$, Algorithm \ref{algorithm1} can reduce the overall computational cost in obtaining optimal polices. Moreover, the computational efficiency of Algorithm \ref{algorithm1} compared to traditional relative value iteration will be shown in the simulation part.

\begin{algorithm}[t]
\caption{Modified Relative Value Iteration}
\hspace*{0.02in}

\label{algorithm1}
\begin{algorithmic}[1]
\State  initialize $V_{0} (s) \leftarrow 0$, $\forall s \in \mathcal{S}$; $\epsilon \leftarrow 0.01 $;  $k \leftarrow 1$;

\State  compute $V_{1} (s) $ and $\pi_1^{*}(s) $ by (\ref{v_function})-(\ref{optimalpi}), $\forall s \in \mathcal{S}$;

\State   {\bf repeat}

\State   {\bf for } $\forall s \in \mathcal{S} $

\State  \quad {\bf if } $ \exists \ s' \in \mathcal{S} $ such that $\pi_{k+1}^{*}  (s') = \bar a \in \mathcal{A} $ and $\bar a_i = 1$

\State   \qquad set $V_{k+1}(s)$ and $\pi_{k+1}^{*}  (s)$ for all $ s' \leq_i s$ as

  $ V_{k+1}(s) \leftarrow    c(s',\bar a) + \sum_{s^{+} \in \mathcal{S} } V_k (s^{+}) \mathcal{P}(s^{+}|s',\bar a)$;

  $\pi_{k+1}^{*}  (s)  \leftarrow \bar a$;

\State     \quad {\bf else }

\State    \qquad compute $V_{k+1}(s)$ and $\pi_{k+1}^{*}  (s)$ by (\ref{v_function})-(\ref{optimalpi});

\State   \quad {\bf end if }

\State  \quad $V_{k+1}(s) \leftarrow V_{k+1}(s) -   V_{k+1}(s_0)$, $s_0 $ is a fixed state;

\State   {\bf end for }

\State    $ k  \leftarrow k + 1$;

\State    {\bf until} $\| V_{k}(s) - V_{k-1}(s))\|    \leq \epsilon$

\end{algorithmic}

{\bf Output: $\pi^{*}(s) $, $\forall s \in \mathcal{S}$;}
\end{algorithm}

%

The optimal scheduling policies for two special cases are provided in the following corollaries.

\begin{assumption} \label{specialcase}
$Q_i$ and $R_i$ are identical for sensors, and the communication graph with $\mathcal{A}=[a_{ij}]_{N \times N}$ is strongly connected.
\end{assumption}

\begin{corollary} \label{corollary1}
If Assumptions \ref{feasibilitycondition} and \ref{specialcase} hold, an optimal policy $\pi^{*}$ is
\begin{align}
 & \pi^{*}_i (s) =    \begin{cases}
   1,    & \text{if $ s_{i} \in \max_M \{ s \}  $ }, \\
   0,    & \text{if $ s_{i} \notin \max_M \{ s \}  $ }, \\
  \end{cases}  \notag
\end{align}
$\forall i=1,\ldots,N$, where $\max_M \{ s \} $ denotes the largest $M$ elements in $s$.
\end{corollary}

The proof of Corollary \ref{corollary1} is similar to that for Lemma \ref{lemmathm2} and Theorem \ref{thm2}. Corollary \ref{corollary1} presents an intuitive result, i.e.,  for coupled identical systems, the optimal policy is to perform scheduling for sensors with the most number of successive packet drops. When there exist only two coupled subsystems, an optimal policy with the switching feature is obtained in closed-form as follows.

\begin{corollary} \label{corollary2}
If Assumptions \ref{feasibilitycondition} and \ref{specialcase} hold, $N=2$ and $M=1$,   an optimal policy $\pi^{*}$ is
\begin{align}
 & (\pi^{*}_1 (s),\pi^{*}_2 (s)) =    \begin{cases}
   (1,0),    & \text{if $ \tau_1 > \tau_2  $ }, \\
   (0,1),    & \text{if $ \tau_1 \leq \tau_2  $ }. \\
  \end{cases}  \notag
\end{align}
\end{corollary}

\section{Simulation} \label{simulation}
In this section, several numerical examples are presented to show the effectiveness of the theoretical results. Before moving on, define an average reward to evaluate the scheduling performance as:
\begin{align}
& J_{AVG} \triangleq     \frac{1}{T}   \sum_{k=1}^{T} \sum_{i=1}^{2} \Big ( \text{Tr}(P_{k,i}) +  \kappa \alpha_{k,i} \Big ) . \notag
\end{align}

\begin{figure}[!htb]
\center
\subfigure{{\includegraphics[scale=0.5]{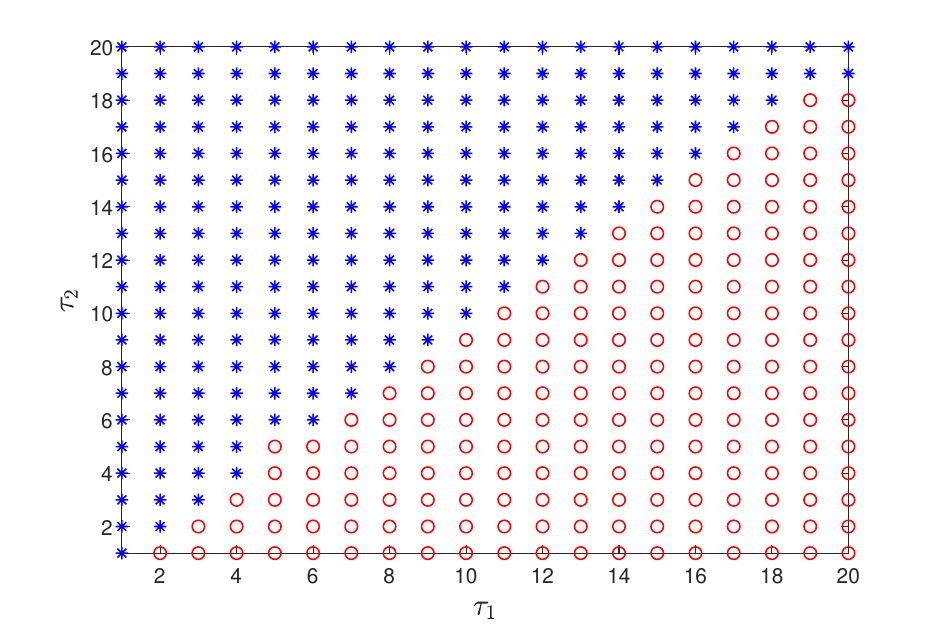}}}
\caption{The visualization of the threshold policy, where $\lambda_1= \lambda_2=0.8$;  both the transmissions of sensors 1 and 2 are subject to one-step time delay;  `{\color{red}o}' denotes the scheduling of sensor 1 and `{\color{blue}*}' denotes the scheduling of sensor 2.}  \label{f:switching6}
\end{figure}

\begin{figure}[!htb]
\center
\subfigure{{\includegraphics[scale=0.5]{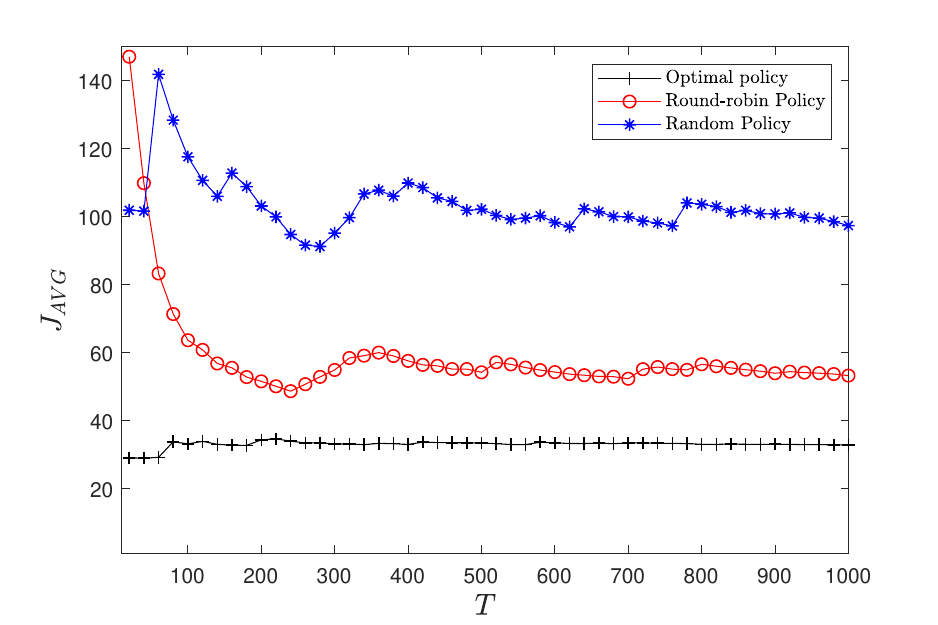}}}
\caption{The estimation performance under different policies. } \label{f:com}
\end{figure}

\textbf{Case 1:}  A complex network with $N=2$ nodes is considered, with dynamics being described by (\ref{system_model}) and parameters being taken from \cite{wu2020Optimal1} as follows:
\begin{align}
  A = \left[{
\begin{array}{*{20}{c}}
  { 1.1} & {0.5}  \\
  { 0} & {0.9}  \\
  \end{array} }\right], \qquad
  C = Q_i = R_i = G = \left[{
\begin{array}{*{20}{c}}
  { 1} & {0}  \\
  { 0} & {1}  \\
  \end{array} }\right],
   \notag
\end{align}
for all $i=1,2$. Besides, the coupling strength is $\mu=0.1$ and the successful transmission rate is  $\lambda_1= \lambda_2=0.8$. Nodes $1$ and $2$ are neighbors, i.e., $a_{12}=a_{21}=1$. The initial states are chosen as $x_{0,1} = [1, \ 1]$ and $x_{0,2} = [2, \ 2]$. The initial state estimates are chosen as $\hat{x}_{0,1}^s = x_{0,1} + w_1$ and $\hat{x}_{0,2}^s = x_{0,2} +w_2$, where $w_1$ and $w_2$ are zero-mean Gaussian noise with covariances $P_{1}$ and $P_{2}$, respectively. For $\textbf{Problem 1} $, the coefficient of scheduling cost is $\kappa=20$ and only one sensor is allowed to send its information at each step.

According to Theorem \ref{thm1}, since $(1-\lambda) \rho^{2r} =  0.79 <1$, $\textbf{Problem 1} $ is feasible. Besides, let the truncated states $\tau_i \leq 50$, $i=1, \ 2$. The initial value function $V_0(s)$ and the accuracy parameter $\epsilon$ in Algorithm \ref{algorithm1} are set as $0$ and $0.01$, respectively. The simulation results are demonstrated from two aspects: 1) the visualization of the threshold policy; 2) a scheduling performance comparison with two typical methods.

First, as shown in Fig.~\ref{f:switching6}, the optimal policy with the threshold structure is presented. Note that there exists a curve between $\tau_1$ and $\tau_2$. On both sides of the curve, when the number of the corresponding successive packet drops increases, the sensor scheduling policy still remains valid, which verifies the results stated in Theorem \ref{thm2}. Then, a comparative simulation with the round-robin scheduling policy and the random scheduling policy is presented. As shown in Fig.~\ref{f:com}, the average reward under the policy presented in this paper is the minimum, which means that this policy outperforms other typical policies.

The most computationally expensive step in the above process is to obtain the optimal policy by solving {\bf Problem 1}. To reduce the computational cost, a modified method using the threshold structure is proposed  as summarized  in  Algorithm \ref{algorithm1}. The computational time of  Algorithm \ref{algorithm1} and the traditional brute-force relative value iteration method \cite{zhou2017optimal} are compared by MATLAB on a Windows 64-Bit server with Intel(R) Core(TM) i7-10710U CPU. Specifically, the computational time of Algorithm \ref{algorithm1} is $2.05s$ while the one of the traditional method is $3.53s$. Moreover, the iteration number of step $8$ in Algorithm 1  is $22033$ while the one in the traditional method is $46550$. The numerical result demonstrates that the proposed method  possesses higher computational efficiency.


\begin{figure}[!htb]
\center
\subfigure{{\includegraphics[scale=0.5]{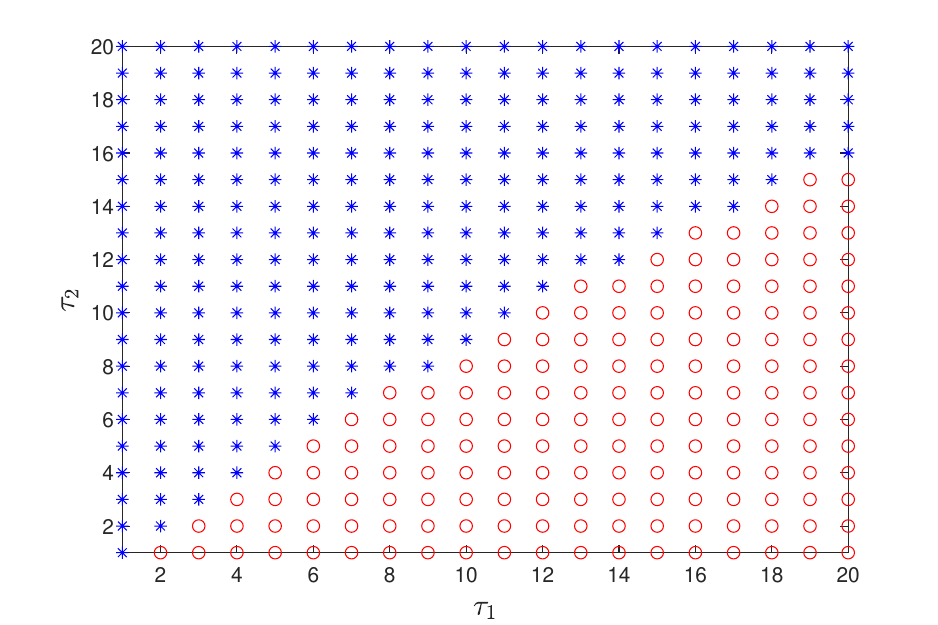}}}
\caption{The visualization of the threshold policy, where $\lambda_1=1$,  $\lambda_2=0.8$; both the transmissions of sensors 1 and 2 are subject to one-step time delay; `{\color{red}o}' and `{\color{blue}*}' are defined in Fig. \ref{f:switching6}.} \label{f:packetloss6}
\end{figure}

\begin{figure}[!htb]
\center
\subfigure{{\includegraphics[scale=0.5]{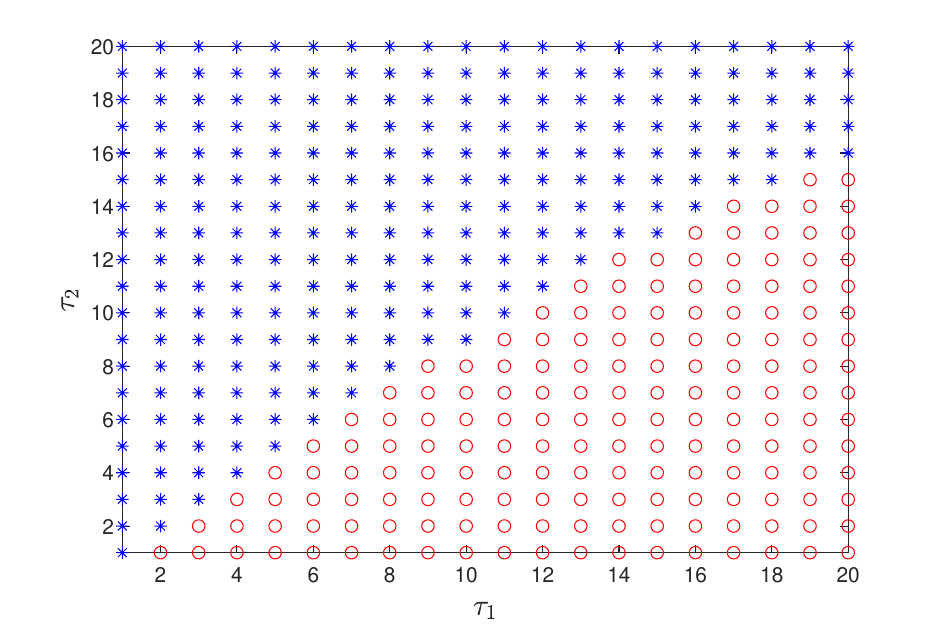}}}
\caption{The visualization of the threshold policy, where $\lambda_1 = \lambda_2 = 0.8$; only the transmission of sensor 2 is subject to one-step time delay;  `{\color{red}o}' and `{\color{blue}*}' are defined in Fig. \ref{f:switching6}.} \label{f:policy_time}
\end{figure}

\textbf{Case 2:} The effects of uncertain processes, including packet loss and one-step time delay, on the scheduling policy are illustrated. First, consider the effect of packet loss. Noting that the packet-loss phenomenon studied in this paper is represented by the successful transmission rate of sensors, i.e., $\lambda_i$ in Section \ref{sec2.3}. Here, assume that there is no packet loss in the transmission from sensor $1$ to node $1$, i.e., $\lambda_1 = 1$, and other settings remain the same as in \textbf{Case 1}. The optimal policy in this situation is shown in  Fig.~\ref{f:packetloss6}. By comparing the scheduling sensors in Figs. \ref{f:switching6} and \ref{f:packetloss6} at the same state $(\tau_1, \ \tau_2)$, it can be seen that the threshold structure still exists in Fig. \ref{f:packetloss6} but the switching curve inclines towards the sensor with a higher successful transmission rate. This indicates that a lower successful transmission rate leads to a higher transmission frequency or authority to guarantee the overall performance. Similarly, consider the situation where no time delay exists in the transmission from sensor $1$ to node $1$ and other settings remain the same as in \textbf{Case 1}. The optimal policy is shown in Fig.~\ref{f:policy_time}, where the switching curve inclines towards the sensor without transmission delay. This indicates that the time-delay transmission needs a higher transmission authority to guarantee the overall performance. In conclusion, the uncertain processes do not determine the existence of the threshold structure but they have an effect on the threshold.

\begin{figure}[!htb]
\center
\subfigure{{\includegraphics[scale=0.5]{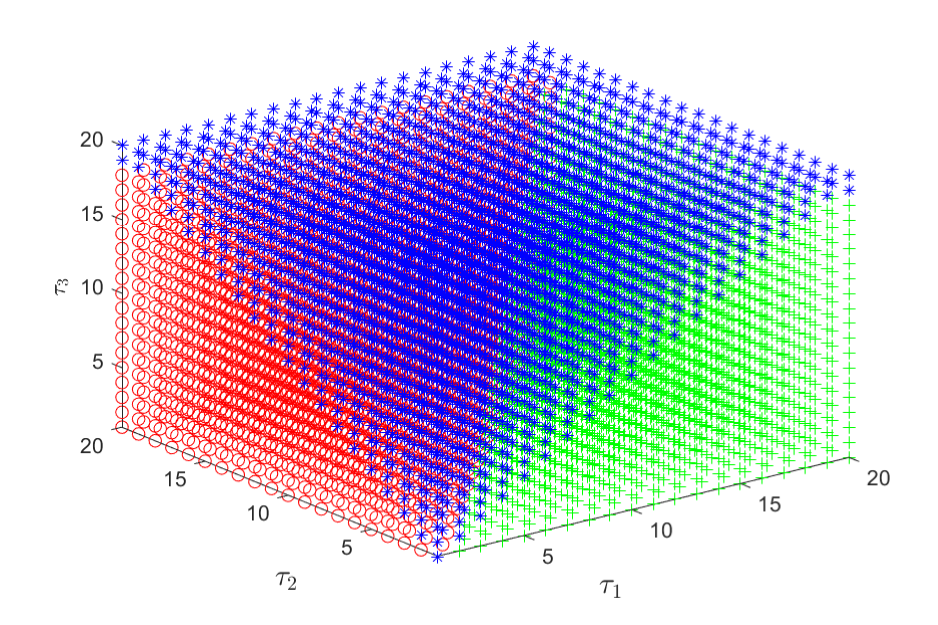}}}
\caption{The visualization of actions, where `{\color{green}+}' denotes the scheduling of sensor 1, `{\color{red}o}' denotes the scheduling of sensor 2 and `{\color{blue}*}' denotes the scheduling of sensor 3. } \label{f:3_D}
\end{figure}

\textbf{Case 3:} To better illustrate the theoretical results obtained in Theorem \ref{thm2}, the policy with a 3-dimensional display is depicted, where the switching curves between sensors are hypersurfaces. In this simulation, a network with $N = 3$ nodes is considered. The system parameters remain valid with the coupling $a_{1,2}=a_{2,1}=1$, $a_{1,3}=a_{3,1}=1$, and $a_{2,3}=a_{3,2}=1$, and the successful transmission rates $\lambda_1 = \lambda_2= \lambda_3=0.9$. As shown in Fig.~\ref{f:3_D}, actions are split by several hypersurfaces, which is consistent with the result in Theorem \ref{thm2}.

\section{Conclusion} \label{Conclusion}
In this paper, optimal sensor scheduling for a complex network  was considered, where each node was equipped with a remote sensor. All sensors shared a common and unstable information transmission channel, where only a portion of them could send the estimates to the associated nodes with one-step time delay. After the node received the estimates, it processed them according to the prior-estimator. Based on a distributed state estimation framework,  an average sum of estimation error bounds over the infinite horizon was regarded as the  index for evaluating the scheduling performance. First, this problem was formulated as a standard MDP problem. Then, the existence of a deterministic and stationary optimal policy was ensured by a sufficient condition. Moreover, it was shown that the optimal policy had a threshold structure, resulting in feasible online implementation. In the end, numerical examples were shown to illustrate the theoretical results.

\section*{Appendices}
\appendix

\section{Proof of Lemma \ref{lemma2}} \label{lemma2proof}
First, it follows from (\ref{system_model})-(\ref{posterioriestimate}) that the estimation error $e_{k+1,i}$ can be derived as
\begin{align}
e_{k+1,i}  = & (I_n - K_{k+1,i}C)\Big (A e_{k,i} + \mu \sum_{j=1}^{N} a_{ij} G e_{k,j}  - \omega_{k,i} \Big )   \notag \\
   & + K_{k+1,i} \nu_{k+1,i}. \notag
\end{align}
Hence,
\begin{align}
& \mathbb{E} \{ e_{k+1,i} \} \notag \\
 = & (I_n - K_{k+1,i}C)\Big (A \mathbb{E} \{ e_{k,i} \} + \mu \sum_{j=1}^{N} a_{ij} G \mathbb{E} \{ e_{k,j} \}  \Big ).   \notag
\end{align}
Further, the expected value of $e_{k+1,i}$ satisfies $\mathbb{E} \{ e_{k+1,i} \} = \mathbb{E} \{ e_{k,i} \} = \cdots = \mathbb{E} \{ e_{0,i} \} = 0$, which indicates that the proposed estimator is unbiased. By using Young's inequality, the estimation error covariance of node $i$ at step $k+1$, i.e., $\mathbb{E} \{e_{k+1,i}e_{k+1,i}^T \}$, satisfies
\begin{align}
 & \mathbb{E} \{e_{k+1,i}e_{k+1,i}^T \} \leq  (I_n - K_{k+1,i}C) \Big [(1+\mu)A \mathbb{E} \{e_{k,i}e_{k,i}^T \}  A^T  \notag \\
&   + (\mu + \mu^2) d_{i} \sum_{j \in \mathcal{N}_i}  G \mathbb{E} \{e_{k,j}e_{k,j}^T \}  G^T + Q_i \Big ]  (I - K_{k+1,i}C)^T   \notag \\
&   + K_{k+1,i} R_i K_{k+1,i}^T. \notag
\end{align}
Then, define two auxiliary matrix functions as follows:
\begin{align}
 P_{k+1,i}^d  = & (I_n - K_{k+1,i}C) \breve{P}_{k+1,i}^d   (*)^T  + K_{k+1,i} R_i K_{k+1,i}^T,  \notag \\
  \breve{P}_{k+1,i}^d = & (1+\mu)A P_{k,i}^d   A^T   + (\mu + \mu^2) d_{i} \sum_{j \in \mathcal{N}_i}  G P_{k,j}^d   G^T  + Q_i. \notag
\end{align}
According to \cite[Lemma 4]{peihu2020}, when $P_{0,i}^d \ge  \mathbb{E} \{e_{0,i}e_{0,i}^T \}$, one has $P_{k,i}^d \ge  \mathbb{E} \{e_{k,i}e_{k,i}^T \}$ for all $k\ge 1$. As a result, the gain matrix can be derived by minimizing the trace of this upper bound matrix. Hence,  $K_{k+1,i}$ is obtained as (\ref{Kdis}). Moreover, the evolution of $P_{k,i}^d$ is obtained as (\ref{pd}).

In the following, the uniform boundedness of $P_{k,i}^d$ is  proved. Since $(\tilde{A}, \  d_{m}  \tilde{G}, \ C)$ is detectable, there exists a gain matrix $K$ and a positive definite matrix $P$ such that
\begin{align} \label{PPP}
P = &  (I - KC) (\tilde{A} P \tilde{A}^T )  (*)^T    + d_{m}^2 (I - KC)  \tilde{G} P  \tilde{G}^T   (*)^T  \notag \\
   &  +  (I - KC) Q_m  (*)^T  + K R_m K^T ,
\end{align}
where $R_{m}$ and $Q_{m}$ are defined in Lemma \ref{lemma2}. Then, it suffices to prove that $P_{k,i}^d \leq P$,  $\forall k = 1, \ \ldots$, $\forall i = 1, \ \ldots$, $N$.
By choosing $P_{0,i}^d = 0$, when $P_{k,i}^d \leq P$, $\forall i = 1, \ \ldots$, $N$, it follows from (\ref{Kdis})-(\ref{brevepd}) that
\begin{align} \label{derivationpd}
 & P_{k+1,i}^d \notag \\
  = & ( (\breve{P}_{k+1,i}^d)^{-1}  + C^T R_{i}^{-1} C )^{-1}       \notag  \\
 \overset{\text{a}}{=} & \breve{P}_{k+1,i}^d - \breve{P}_{k+1,i}^d C^T (C \breve{P}_{k+1,i}^d C^T + R_i)^{-1} C  (\breve{P}_{k+1,i}^d)^T \notag \\
 & + (K_{k+1,i} -  \breve{P}_{k+1,i}^d C^T (C \breve{P}_{k+1,i}^d C^T + R_i)^{-1}) \notag \\
 & \times  (C \breve{P}_{k+1,i}^d C^T + R_i)   (*)^{T}   \notag \\
 \leq & \breve{P}_{k+1,i}^d - \breve{P}_{k+1,i}^d C^T (C \breve{P}_{k+1,i}^d C^T + R_i)^{-1} C  (\breve{P}_{k+1,i}^d)^T \notag \\
 & + (K -  \breve{P}_{k+1,i}^d C^T (C \breve{P}_{k+1,i}^d C^T + R_i)^{-1}) \notag \\
 & \times  (C \breve{P}_{k+1,i}^d C^T + R_i) (*)^{T}  \notag \\
 = & (I_n - KC) \breve{P}_{k+1,i}^d  (*)^{T}
  + K  R_i K^T  \notag \\
 \leq &  (I_n - KC) (\tilde{A} P \tilde{A}^T + d_m^2 \tilde{G} P  \tilde{G}^T + Q_m  )  (*)^{T} + K  R_m K^T  \notag \\
 = & P,
\end{align}
where the result of $\overset{\text{a}}{=}$ is obtained using $K_{k+1,i} = \breve{P}_{k+1,i}^d C^T (C \breve{P}_{k+1,i}^d C^T + R_i)^{-1}$. By mathematical induction, $P_{k,i}^d$ is uniformly upper bounded by $P$. Moreover, since $P_{k,i}^d$ is uniformly lower bounded by $0_{n \times n}$, one has that $P_{k,i}^d$ is uniformly bounded.

Next, that $P_{k,i}^d$ is monotonically increasing with respect to $k$ by mathematical induction is proved. First, choose the initial parameter $P_{0,i}^d =  0$, $\forall i=1, \ \ldots, \ N$. Since the initial state $x_{0,i}$ is known as $\overline{x}_{0,i}$, $P_{0,i}^d \ge \mathbb{E} \{(x_{0,i} - \overline{x}_{0,i})(x_{0,i} - \overline{x}_{0,i})^T \} = 0$. Hence, $P_{0,i}^d$ is an upper bound of $ \mathbb{E} \{ e_{0,i} e_{0,i}^T \}$. Also, it follows from (\ref{pd}) and (\ref{brevepd}) that $\breve{P}_{0,i}^d  =   Q_i$ and $ P_{1,i}^d  =  (  Q_i^{-1}  + C^T R_{i}^{-1} C )^{-1} > P_{0,i}^d = 0$. Then, for $P_{k,i}^d > P_{k-1,i}^d $, $\forall i=1, \ \ldots, \ N$, it follows from (\ref{brevepd}) that
  \begin{align}
  &  \breve{P}_{k+1,i}^d - \breve{P}_{k,i}^d  \notag \\
   = & \tilde{A} (P_{k,i}^d - P_{k-1,i}^d)   \tilde{A}^T  +  d_i \sum_{j \in \mathcal{N}_i} \tilde{G} ( P_{k,j}^d  -P_{k-1,j}^d) \tilde{G}^T
      >  0,  \notag
 \end{align}
which indicates that $P_{k+1,i}^d > P_{k,i}^d $, $\forall i=1, \ \ldots, \ N$. Further, according to (\ref{pd}), one has
\begin{align}
& (P_{k+1,i}^d  )^{-1} - (P_{k,i}^d   )^{-1}
=     (\breve{P}_{k+1,i}^d)^{-1}  -  (\breve{P}_{k,i}^d)^{-1} < 0. \notag
\end{align}
This guarantees that  $  \breve{P}_{k+1,i}^d  > \breve{P}_{k,i}^d $. As a result, $P_{k,i}^d$ is monotonically increasing.

Altogether, one can conclude that $P_{k,i}^d$ converges to a constant matrix. In addition, according to (\ref{derivationpd}), the convergence rate of $P_{k,i}^d$ with $K_{k,i}$ is not slower than that with $K$, while the latter is exponentially fast.

\section{Proof of Lemma \ref{lemma4}} \label{lemma4proof}
It suffices to prove that $P_{k,i}^d \leq \Theta$ holds when the condition (\ref{condition}) is satisfied. First, at step $k=0$,  $P_{0,i}^d = 0 \leq \Theta$ holds. Then, if $P_{k,i}^d \leq \Theta$, $\forall i = 1, \ \ldots$, $N$, holds, one has
\begin{align}
   P_{k+1,i}^d
   \leq   &  (I_n - L C) (\tilde{A}  P_{k,i}^d  \tilde{A}^T   + Q_i )  (I_n - L C)^T  + L R_i L^T  \notag \\
 &   +  (I_n - L C) \Big (    \sum_{j \in \mathcal{N}_i} d_{i} \tilde{G}  P_{k,j}^d   \tilde{G}^T \Big )  (I_n - L C)^T  \notag \\
 \leq &   (I_n - L C) ( \tilde{A} \Theta \tilde{A}^T  + Q_m  )  (I_n - L C)^T  + L R_m L^T \notag \\
  & + d_m^2 (I_n - L C)   \tilde{G} \Theta   \tilde{G}^T (I_n - L C)^T   \notag \\
  \overset{\text{b}}{=} & \Theta + (I_n - L C) ( d_m^2 \tilde{G} \Theta  \tilde{G}^T - z I_n ) (I_n - L C)^T \notag \\
  \overset{\text{c}}{\leq} & \Theta, \notag
\end{align}
where the result of $\overset{\text{b}}{=}$ is derived by substituting $\Theta$ defined in Lemma \ref{lemma4} and the result of $\overset{\text{c}}{\leq}$ is derived based on (\ref{condition}). The rest of the proof is similar to that of Lemma \ref{lemma2}, which is omitted.

\section{Proof of Theorem \ref{proposition1}} \label{proposition1proof}
Without loss of generality, it is assumed that elements in $s_{k}$ are ranked in a descending order, i.e., $\tau_{k,1} \ge \tau_{k,2} \ge \cdots \ge \tau_{k,N}$. Then, the time interval $[k-\tau_{k,1}, k]$ is divided into $N$ sub-intervals, i.e., $[k-\tau_{k,1}, k-\tau_{k,2})$, $\ldots$, $[k-\tau_{k,i}, k-\tau_{k,i+1})$, $\ldots$,  $[k-\tau_{k,N}, k]$. In particular, if $\tau_{k,i}=\tau_{k,i+1}$, denote $[k-\tau_{k,i}, k-\tau_{k,i+1})=\varnothing$. By denoting $P_{k}^{sum} = \sum_{i=1}^{N}  P_{k,i} $, the relation between $P_{k}^{sum}$ and $s_{k}$ is revealed next.

First, on the interval $[k-\tau_{k,1}, k-\tau_{k,2})$, it follows from (\ref{Piteration}) that $Y_{k-\tau_{k,1},i} =  P_{i}$, $\forall i=1,\ldots,N$, and  $Y_{l,i} =  P_{i}$, $\forall i=2,\ldots,N$, $\forall l \in [k-\tau_{k,1}, k-\tau_{k,2})$. Hence,  one has 
\begin{align}
&  Y_{k-\tau_{k,2},1} \notag \\
   = & h_1(Y_{k-1-\tau_{k,2},1}, P_{j}, j \in \mathcal{N}_1) \notag \\
   \overset{\text{1}}{=} & \tilde{A} Y_{k-1-\tau_{k,2},1} \tilde{A}^T   +   d_{1} \sum_{j \in \mathcal{N}_1}   \tilde{G}  P_{j}   \tilde{G}^T + Q_1    \notag \\
   = & \tilde{A} h_1(Y_{k-2-\tau_{k,2},1}, P_{j}, j \in \mathcal{N}_1) \tilde{A}^T   +   d_{1} \sum_{j \in \mathcal{N}_1}   \tilde{G}  P_{j}   \tilde{G}^T + Q_1    \notag \\
   \overset{\text{2}}{=} & \tilde{A} ( \tilde{A} Y_{k-2-\tau_{k,2},1} \tilde{A}^T   +   d_{1} \sum_{j \in \mathcal{N}_1}   \tilde{G}  P_{j}   \tilde{G}^T + Q_1  ) \tilde{A}^T   \notag \\
    & +   d_{1} \sum_{j \in \mathcal{N}_1}   \tilde{G}  P_{j}   \tilde{G}^T + Q_1    \notag \\
   = & \tilde{A}^2 Y_{k-2-\tau_{k,2},1} (\tilde{A}^T)^2   +  \sum_{j \in \mathcal{N}_1}    \tilde{A} ( d_{1}  \tilde{G}  P_{j}   \tilde{G}^T + Q_1  ) \tilde{A}^T    \notag \\
   & +  \sum_{j \in \mathcal{N}_1}   d_{1}  \tilde{G}  P_{j}   \tilde{G}^T + Q_1    \notag \\
   \overset{\text{3}}{=} & \tilde{A}^3 Y_{k-3-\tau_{k,2},1} (\tilde{A}^T)^3   +  \sum_{j \in \mathcal{N}_1}    \tilde{A}^2 ( d_{1}  \tilde{G}  P_{j}   \tilde{G}^T + Q_1  ) (\tilde{A}^T)^2 \notag \\
    &  + \sum_{j \in \mathcal{N}_1}    \tilde{A} ( d_{1}  \tilde{G}  P_{j}   \tilde{G}^T + Q_1  ) \tilde{A}^T   +  \sum_{j \in \mathcal{N}_1}   d_{1}  \tilde{G}  P_{j}   \tilde{G}^T + Q_1    \notag \\
   = & \cdots \notag \\
   \overset{\text{4}}{=}  &   \tilde{A}^{\tau_{k,1}-\tau_{k,2}}   P_{1}  (\tilde{A}^T)^{\tau_{k,1}-\tau_{k,2}} \notag \\
   & \qquad +  \sum_{l=0}^{\tau_{k,1}-\tau_{k,2}-1}   \tilde{A}^l  \Big (   d_{1} \sum_{j \in \mathcal{N}_1}   \tilde{G}  P_{j}   \tilde{G}^T + Q_1 \Big  ) (\tilde{A}^l)^T    \notag \\
    \triangleq & \overline{h}_1  (\tau_{k,1},\tau_{k,2}), \notag
\end{align}
where the results of $\overset{\text{1}}{=}$,  $\overset{\text{2}}{=}$ and  $\overset{\text{3}}{=}$ are obtained by substituting the expressions of $ h_1(Y_{k-1-\tau_{k,2},1}$, $P_{j}, j \in \mathcal{N}_1)$, $ h_1(Y_{k-2-\tau_{k,2},1}, P_{j}, j \in \mathcal{N}_1)$ and $  h_1(Y_{k-3-\tau_{k,2},1}, P_{j}$, $j \in \mathcal{N}_1)$, respectively; and the result of $\overset{\text{4}}{=}$ is obtained by operating such process $\tau_{k,1}-\tau_{k,2}$ times.

Then, on the interval $[k-\tau_{k,2}, k-\tau_{k,3})$, it follows from (\ref{Piteration}) that $Y_{k-\tau_{k,2},i} =  P_{i}$, $\forall i=2,\ldots,N$, and  $Y_{l,i} =  P_{i}$, $\forall i=3,\ldots,N$, $\forall l \in [k-\tau_{k,2}, k-\tau_{k,3})$. If nodes $1$ and $2$ are not neighbors, then
\begin{align}
  & Y_{k-\tau_{k,3},1}+Y_{k-\tau_{k,3},2} \notag  \\
  =    &   \sum_{l=0}^{\tau_{k,2}-\tau_{k,3}-1}   \tilde{A}^l  \Big (   d_{1} \sum_{j \in \mathcal{N}_1}   \tilde{G}  P_{j}   \tilde{G}^T + d_{2} \sum_{j \in \mathcal{N}_2}   \tilde{G}  P_{j}   \tilde{G}^T + Q_1 \notag \\
  &   + Q_2 \Big  ) (\tilde{A}^l)^T   + \tilde{A}^{\tau_{k,2}-\tau_{k,3}}   (Y_{k-\tau_{k,2},1}+P_{2}) (\tilde{A}^T)^{\tau_{k,2}-\tau_{k,3}} \notag \\
    \triangleq & \overline{h}_{2}  (\tau_{k,1},\tau_{k,2},\tau_{k,3}) . \notag
\end{align}
Similarly, when nodes $1$ and $2$ are neighbors, $Y_{k-\tau_{k,3},1}+Y_{k-\tau_{k,3},1}$ can still be expressed as a function with variables $\tau_{k,1},\tau_{k,2}$ and $\tau_{k,3}$.
After applying the above derivation technique to intervals $[k-\tau_{k,3}, k-\tau_{k,4})$, $\ldots$,  $[k-\tau_{k,N}, k]$, one has that $P_{k}^{sum}$ is a function of $\tau_{k,i}$, $i =1, \ldots,N$, i.e., there exists a function $\overline{h}$ such that  $P_{k}^{sum} = \overline{h} (\{\tau_{k,i}\}_{i =1}^{N}) = \overline{h} (s_{k})$. Hence, one has
\begin{align}
  & \sum_{i=1}^{N} ( \text{Tr}(P_{k,i}) +  \kappa \alpha_{k,i})
   =   \text{Tr}(P_{k}^{sum}) + \sum_{i=1}^{N} \kappa \alpha_{k,i} \notag \\
  = & \text{Tr}(\overline{h} (s_{k})) + \sum_{i=1}^{N} \kappa \alpha_{k,i}, \notag
\end{align}
which indicates that $ c(s_{0:k},a_{k}) $ is deterministic for a given pair of $(s_{k},a_{k})$. Thus, $ c(s_{0:k},a_{k})$  is independent of $s_{0:k-1}$.

\section{Proof of Theorem \ref{thm1}} \label{thm1proof}
According to \cite[Lemma 3]{wu2020Optimal1} and \cite{hernandez2012discrete}, it suffices to prove that there exists a policy  $\{ \pi_k \}_{k=0}^{\infty}$  such that
\begin{align}
\lim_{T \rightarrow  \infty} \text{sup} \frac{1}{T} \mathbb{E} \bigg [  \sum_{k=1}^{T} \sum_{i=1}^{N} \Big ( \text{Tr}(P_{k,i}) +  \kappa \alpha_{k,i} \Big ) \bigg] < \infty . \notag
\end{align}

To do so, since $\alpha_{k,i} \in \{0, 1\}$ and $N$ is finite, one only needs to prove that there exists a policy  $\{ \pi_k \}_{k=0}^{\infty}$  such that
\begin{align} \label{boundedcondition}
\lim_{T \rightarrow  \infty} \text{sup} \frac{1}{T} \mathbb{E} \bigg [  \sum_{k=1}^{T} \sum_{i=1}^{N}  \text{Tr}(P_{k,i})   \bigg] < \infty .
\end{align}

In the following, it is proved that if Assumption \ref{feasibilitycondition} is satisfied, the above inequality will hold under a modified round-robin scheduling policy.

First, denoting $P_{k}^{sum} = \sum_{i=1}^{N}  P_{k,i} $, $Q^{sum} = \sum_{i=1}^{N}  Q_{i} $ and $Y_{k}^{sum} = \sum_{i=1}^{N}  Y_{k,i} $, it follows from (\ref{Piteration}) that
\begin{align}
P_{k+1}^{sum} = & \sum_{i=1}^{N} \Big ( \tilde{A} Y_{k,i}  \tilde{A}^T  +  d_{i} \sum_{j \in \mathcal{N}_i}  \tilde{G} Y_{k,i}   \tilde{G}^T + Q_i \Big ) \notag \\
= & \tilde{A} Y_{k}^{sum} \tilde{A}^T +  \sum_{i=1}^{N}  \sum_{j \in \mathcal{N}_i}  d_{m} \tilde{G} Y_{k,i}   \tilde{G}^T +  Q^{sum}   \notag \\
\leq & \tilde{A} Y_{k}^{sum} \tilde{A}^T  +  d_m^2    \tilde{G} Y_{k}^{sum}   \tilde{G}^T + Q^{sum}.     \notag
\end{align}
Hence,
\begin{align}
& \sum_{i=1}^{N}  \text{Tr}(P_{k+1,i})  =     \text{Tr} ( P_{k+1}^{sum} )
\notag \\
\leq  & \text{Tr}(\tilde{A} Y_{k}^{sum} \tilde{A}^T   +   d_m^2   \tilde{G} Y_{k}^{sum}   \tilde{G}^T ) + \text{Tr}(Q^{sum}) \notag \\
\overset{\text{d}}{=} & \text{Tr}( Y_{k}^{sum} ( \tilde{A}^T \tilde{A} +   d_m^2     \tilde{G}^T  \tilde{G})) + \text{Tr}(Q^{sum}) \notag \\
=  & \text{Tr}( Y_{k}^{sum} U^T U) + \text{Tr}(Q^{sum}) \notag \\
=  & \text{Tr}( U Y_{k}^{sum} U^T) + \text{Tr}(Q^{sum}), \notag
\end{align}
where the result of $ \overset{\text{d}}{=}$ is derived based on the fact that $ \text{Tr} (AB) = \text{Tr} (BA) $ and $ \text{Tr} (A+B) = \text{Tr} (A) +\text{Tr} (B) $ for any  matrices $A$ and $B$ of appropriate dimensions, and $U$ is defined in Assumption \ref{feasibilitycondition}. Therefore, to ensure (\ref{boundedcondition}), it suffices to prove that
\begin{align}
\lim_{T \rightarrow  \infty} \text{sup} \frac{1}{T} \mathbb{E} \bigg [  \sum_{k=1}^{T} \text{Tr}( Y_{k}^{sum})   \bigg] < \infty . \notag
\end{align}

\begin{figure}[!htb]
\center
\subfigure{{\includegraphics[scale=0.3]{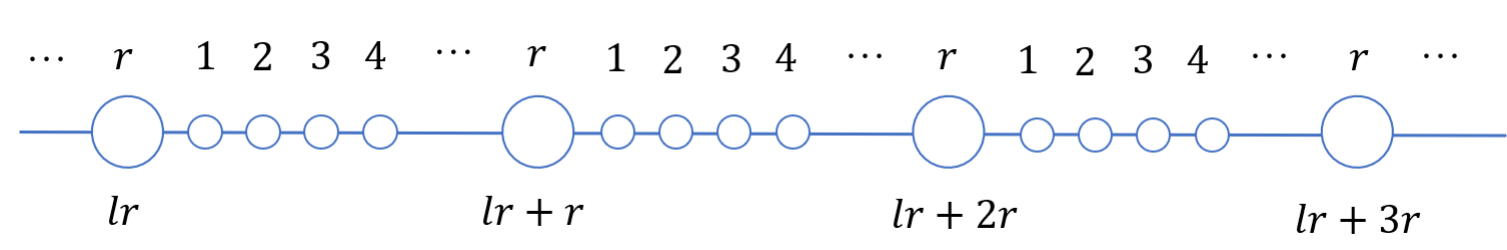}}}
\caption{The time line.} \label{f:round_robin}
\end{figure}

Now, a modified round-robin scheduling policy is presented as follows. As shown in Fig.~\ref{f:round_robin},  the time line is divided into a sequence of time intervals, where $r=\lceil\frac{N}{M}\rceil$. During each time interval $(lr,lr+r]$, $\forall l=0,1,\ldots$, each sensor is allowed to send information to the associated node  once in a fixed round-robin order. Specifically, $N$ sensors are divided into $r$ groups, i.e., $\{(1,\ldots,M),\ldots,(rM-2M+1,\ldots,rM-M),(rM-M+1,\ldots,N)\} \triangleq \{ \mathcal{G}_h \}_{h=1}^{r}$. Sensors in the $h$-th groups are allowed to transmit information at time $lr+h$, $h=1,\ldots,r$. Equivalently, when $h-1 < \lceil\frac{i}{M}\rceil\leq  h $, sensor $i$ is allowed to transmit its information packet at time $lr+h$, $h=1,\ldots,r$.

In the following, the relation between $Y_{lr+r}^{sum}$ and $Y_{lr}^{sum}$ is revealed. First, based on the above modified round-robin scheduling policy, when the current time $k = lr+r$, it follows from (\ref{Piteration}) that
\begin{align}
Y_{lr+1}^{sum}
= &  \sum_{i =1}^{N}  (1-\gamma_{lr+1,i})\Big ( \tilde{A} Y_{lr,i}  \tilde{A}^T  +  d_{i} \sum_{j \in \mathcal{N}_i}  \tilde{G} Y_{lr,j}   \tilde{G}^T   \Big ) \notag \\
&  + \sum_{i =1}^{N}  (1-\gamma_{lr+1,i}) Q_i   +  \sum_{i =1}^{N} \gamma_{lr+r,i}  P_i  . \notag
\end{align}
Further,  $\mathbb{E} \{ \text{Tr}(Y_{lr+1}^{sum}) \}$ is derived as follows:
\begin{align}
& \mathbb{E} \{ \text{Tr}(Y_{lr+1}^{sum}) \} \notag \\
= &  (1-\lambda) \sum_{i =1}^{N} \text{Tr} \Big ( \tilde{A} \mathbb{E} \{ Y_{lr,i}\}  \tilde{A}^T  +  d_{i} \sum_{j \in \mathcal{N}_i}  \tilde{G} \mathbb{E} \{  Y_{lr,j}\}   \tilde{G}^T   \Big ) \notag \\
&  +   \text{Tr} (\hat{Q}_1)  \notag \\
\leq   &  (1-\lambda)  \text{Tr} ( \tilde{A} \mathbb{E} \{ Y_{lr}^{sum} \}  \tilde{A}^T  +  d_{m}^2   \tilde{G} \mathbb{E} \{  Y_{lr}^{sum} \}   \tilde{G}^T     )   + \text{Tr} (\hat{Q}_1)  \notag \\
=   &  (1-\lambda)  \text{Tr}  ( U \mathbb{E} \{ Y_{lr}^{sum} \}  U^T    )   +  \text{Tr} (\hat{Q}_1)  \notag \\
\overset{\text{e}}{\leq} & (1-\lambda) \rho^2 \mathbb{E} \{ \text{Tr}(Y_{lr}^{sum}) \}  +  \text{Tr} (\hat{Q}_1),  \notag
\end{align}
where the result of $ \overset{\text{e}}{\leq}$ is derived by  $\rho = \rho(U)$ defined in Assumption \ref{feasibilitycondition}, and
\begin{align}
& \hat{Q}_1 \triangleq \sum_{i =1}^{N}   (Q_i-\lambda   Q_i   +  \lambda P_i).  \notag
\end{align}

For $Y_{lr+2}^{sum}$,  one has
\begin{align}
&  Y_{lr+2}^{sum} \notag \\
= &  \sum_{i = 1}^{N}   (1-\gamma_{lr+2,i}) \Big ( \tilde{A} Y_{lr+1,i}  \tilde{A}^T  +  d_{i} \sum_{j \in \mathcal{N}_i}  \tilde{G} Y_{lr+1,j}   \tilde{G}^T   \Big ) \notag \\
&  + \sum_{i \notin \mathcal{G}_1}  (1-\gamma_{lr+2,i}) Q_i   +  \sum_{i \notin \mathcal{G}_1}  \gamma_{lr+2,i}  P_i   \notag \\
& + \sum_{i \in \mathcal{G}_1} \gamma_{lr+2,i} \Big ( \tilde{A} Y_{lr+1,i}  \tilde{A}^T  +  d_{i} \sum_{j \in \mathcal{N}_i}  \tilde{G} Y_{lr+1,j}   \tilde{G}^T   \Big ) \notag \\
\overset{\text{f}}{\leq}    &  \sum_{i = 1}^{N}  \Big ( \tilde{A} Y_{lr+1,i}  \tilde{A}^T  +  d_{i} \sum_{j \in \mathcal{N}_i}  \tilde{G} Y_{lr+1,j}   \tilde{G}^T   \Big ) \notag \\
&  + \sum_{i \notin \mathcal{G}_1}  (1-\gamma_{lr+2,i}) Q_i   +  \sum_{i \notin \mathcal{G}_1}  \gamma_{lr+2,i}  P_i ,  \notag
\end{align}
where the result of $\overset{\text{f}}{\leq}$ is derived based on the fact that $\gamma_{lr+2,i}=0$, $\forall i \in \mathcal{G}_1$. Hence, the relation between $\mathbb{E} \{ \text{Tr}(Y_{lr+1}^{sum}) \}$ and $\mathbb{E} \{ \text{Tr}(Y_{lr+1}^{sum}) \}$ can be established as
\begin{align}
 \mathbb{E} \{ \text{Tr}(Y_{lr+2}^{sum}) \}
 {\leq} & \rho^2 \mathbb{E} \{ \text{Tr}(Y_{lr+1}^{sum}) \}  +  \text{Tr} (\hat{Q}_2),   \notag
\end{align}
where
\begin{align}
& \hat{Q}_2 = \sum_{i \notin \mathcal{G}_1}   (Q_i-\lambda   Q_i   +  \lambda P_i).  \notag
\end{align}

Then, after some computations, one obtains
\begin{align}
 & \mathbb{E} \{ \text{Tr}(Y_{lr+r}^{sum}) \} \notag \\
 \leq & \rho^2 \mathbb{E} \{ \text{Tr}(Y_{lr+r-1}^{sum}) \}  +  \text{Tr} (\hat{Q}_r)   \notag \\
 \leq & \rho^4 \mathbb{E} \{ \text{Tr}(Y_{lr+r-2}^{sum}) \}  +  \rho^2 \text{Tr} (\hat{Q}_{r-1})  +  \text{Tr} (\hat{Q}_r) \notag \\
\leq & \cdots \notag \\
\leq & m( \mathbb{E} \{ \text{Tr}(Y_{lr}^{sum}) \}) ,  \notag
\end{align}
where $\hat{Q}_i$, $i=1,\ldots,N$, are constant matrices, and
\begin{align}
m(X)  =   (1-r)\rho^{2r} X + \sum_{h=0}^{r-1} \rho^{2h} \text{Tr} (\hat{Q}_{r-h}).  \notag
\end{align}
Note that, at the current time $lr+r$, $Y_{lr}^{sum}$ is re-updated and thus smaller than the one updated at the time $lr$. Hence,  $\mathbb{E} \{ \text{Tr}(Y_{lr+r}^{sum}) \}$ updated at time $lr+r$ is smaller than $m( \mathbb{E} \{ \text{Tr}(Y_{lr}^{sum}) \})$ updated at time $lr$. According to  \cite{zhou1996robust}, when Assumption \ref{feasibilitycondition} holds and $\hat{Q}_i>0$, $X = m(X)$ has a unique strictly positive definite solution, which indicates that $\mathbb{E} \{ \text{Tr}(Y_{lr}^{sum}) \}$ is uniformly upper bounded, i.e., there exists a positive constant $\kappa$ such that $\mathbb{E} \{ \text{Tr}(Y_{lr}^{sum}) \} \leq \kappa $,  $ \forall l = 0, \ 1, \ \ldots$. Further, based on the fact that $r$ is a finite integer, one can conclude that $\mathbb{E} \{ \text{Tr}(Y_{k}^{sum}) \}$, $ \forall k = 0, \ 1, \ \ldots$, is uniformly upper bounded. Consequently, one has
\begin{align}
\lim_{T \rightarrow  \infty} \text{sup} \frac{1}{T} \mathbb{E} \bigg [  \sum_{k=1}^{T} \text{Tr}( Y_{k}^{sum})   \bigg] < \kappa < \infty . \notag
\end{align}
Thus, the proof of Theorem \ref{thm1} is complete.

\section{Proof of Lemma \ref{lemmathm2}} \label{lemmathm2proof}
Before proceeding, all nodes are divided into $\tau_i$ groups, i.e., $\mathcal{J}_{\tau_i} \triangleq \{h|\tau_h \ge \tau_i \}$ and  $\mathcal{J}_{h} \triangleq \{l|\tau_l =  h \}$, $h=1,\ldots,\tau_i-1$. Then, since $Z_{k - \tau_{i}}^{0} \leq_i  Z_{k - \tau_{i}}^{1}$, one has $Z_{k - \tau_{i},i}^{0} \leq  Z_{k - \tau_{i},i}^{1}$ and $Z_{k - \tau_{i},j}^{0} =  Z_{k - \tau_{i},j}^{1}$, $\forall j \neq i$, which yields
\begin{align}
 & Z_{k+1 - \tau_{i},h}^{0} \notag \\
 =&  \tilde{A} Z_{k- \tau_{i},h}^{0}  \tilde{A}^T   +   d_{h} \sum_{j \in \mathcal{N}_h}  \tilde{G} Z_{k- \tau_{i},j}^{0}   \tilde{G}^T + Q_h      \notag \\
 =&  \tilde{A} Z_{k - \tau_{i},h}^{0}  \tilde{A}^T   +   d_{h} \sum_{j \in \mathcal{N}_h \cap \mathcal{J}_{\tau_i}  }  \tilde{G} Z_{k- \tau_{i},j}^{0}   \tilde{G}^T + Q_h      \notag \\
 & +  d_{h} \sum_{j \in \mathcal{N}_h / \mathcal{J}_{\tau_i}  }  \tilde{G} P_{j}   \tilde{G}^T \notag \\
 \leq &  \tilde{A} Z_{k - \tau_{i},h}^{1}  \tilde{A}^T   +   d_{h} \sum_{j \in \mathcal{N}_h \cap \mathcal{J}_{\tau_i}  }  \tilde{G} Z_{k- \tau_{i},j}^{1}   \tilde{G}^T + Q_h      \notag \\
  & +  d_{h} \sum_{j \in \mathcal{N}_h / \mathcal{J}_{\tau_i}  }  \tilde{G} P_{j}   \tilde{G}^T \notag \\
  =&  \tilde{A} Z_{k- \tau_{i},h}^{1}  \tilde{A}^T   +   d_{h} \sum_{j \in \mathcal{N}_h}  \tilde{G} Z_{k- \tau_{i},j}^{1}   \tilde{G}^T + Q_h      \notag \\
  = & Z_{k+1 - \tau_{i},h}^{1},  \notag
\end{align}
for all $h \in \mathcal{J}_{\tau_i}$, and $Z_{k+1 - \tau_{i},h}^{0}= Z_{k+1 - \tau_{i},h}^{1} = P_{h}$, $\forall h \notin \mathcal{J}_{\tau_i}$.
By mathematical induction, one has $Z_{l,h}^{0} \leq Z_{l,h}^{1}, h \in \mathcal{J}_{\tau_i} \cup \cdots \cup  \mathcal{J}_{k+1-l}$, $\forall l = k+2-\tau_i, \ldots,k $. Besides, $Z_{l,h}^{0}= Z_{l,h}^{1} = P_{h}$, $\forall h \in \mathcal{J}_{1} \cup \cdots \cup  \mathcal{J}_{k-l}$, $\forall l = k+2-\tau_i, \ldots,k $. Hence, $Z_{k,h}^{0} \leq Z_{k,h}^{1}$, $\forall h = 1, \ldots, N$.

\section{Proof of Theorem \ref{thm2}} \label{thm2proof}
Before moving on, two definitions are introduced.

\begin{definition} \label{monotonic}
A measurable function $q(s,a)$ is said to be \textit{monotonic} if $q(s,a) \leq q(s',a)$, $\forall s \leq_i s'$.
\end{definition}

\begin{definition} \label{submodular}
A measurable function $q(s,a)$ is said to be \textit{submodular} if $q(s,a) + q(s',a')  \leq q(s',a) + q(s,a') $, $\forall s \leq_i s', \ a \leq_i a'$.
\end{definition}

Then, according to \cite{ren2018attack}, it suffices to prove that the value function $V(\cdot)$ is \textit{monotonic} and \textit{submodular} on $\mathcal{S}$.

\textbf{Part 1: (\textit{Monotonic})}
First, a discounted average reward is introduced as
\begin{align}
 & \hat{J}_{\delta}(s_0,\pi)  \triangleq    \mathbb{E}_{s_0}^{\pi} \bigg [  \sum_{k=1}^{T}  \delta^k c(s_{k},a_{k}) \bigg], \notag
\end{align}
where $\delta \in (0,1)$ is the discounted factor. Then, a discounted cost MDP is formulated as
\begin{align}
 &   \min_{ \pi } \hat{J}_{\delta}(s_0,\pi). \notag
\end{align}
Based on the Bellman optimality equation, the value function for the above MDP under an optimal policy satisfies
\begin{align} \label{bellmanopt}
 & \hat{V}_{k}(s) =  \min_{ a \in  \mathcal{A} } \Big [ c(s,a) +  \delta \sum_{s^{+} \in \mathcal{S} } \hat{V}_{k+1}(s^{+}) \mathcal{P}(s^{+}|s,a) \Big ] . \notag
\end{align}
Then, according to the vanishing discount approach \cite{hernandez2012discrete}, to guarantee the monotonicity of $V(\cdot)$, it suffices to show that $\hat{V}(\cdot)$ is \textit{monotonic} on $\mathcal{S}$. By the induction-based method, it suffices to guarantee that $c(s,a)$ and $\sum_{s^{+} \in \mathcal{S} } \mathcal{P}(s^{+}|s,a)$ are \textit{monotonic}.

For the former, it suffices to prove that $ Y_{k-1,h}^{s}  \leq  Y_{k-1,h}^{s'} $, $\forall s \leq_i s'$, $\forall h=1,\ldots,N$, where $Y_{k-1,h}^{s}$ and $ Y_{k-1,h}^{s'}$ are defined in (\ref{Piteration}), and the superscripts $s$ and $s'$ denote different MDP states. First, consider the case of $s'_i-s_i=1$. When the current step is $k$, it follows from (\ref{Piteration}) that
\begin{align}
  & Y_{k  - s_{i},i}^{s'}
 =   \tilde{A} Y_{k - s'_{i},i}^{s'}  \tilde{A}^T   +   d_{i} \sum_{j \in \mathcal{N}_i}  \tilde{G} Y_{k - s'_{i},j}^{s'}   \tilde{G}^T + Q_i     \notag \\
\ge &  \tilde{A} P_i  \tilde{A}^T   +   d_{i} \sum_{j \in \mathcal{N}_i}  \tilde{G} P_j   \tilde{G}^T + Q_i
\ge   P_i = Y_{k  - s_{i},i}^{s}. \notag
\end{align}
Besides, one has $Y_{k  - s_{i},h}^{s'}   = Y_{k  - s_{i},h}^{s}$,  $ \forall h \neq i$. Based on the above inequality and this equation, one has $Y_{k  - s_{i}}^{s}  \leq_i Y_{k  - s_{i}}^{s'} $. Then, by using Lemma \ref{lemmathm2}, one can conclude that $Y_{k,h}^{s} \leq Y_{k,h}^{s'}$, $\forall s \leq_i s'$,  i.e., $c(s,a)$ is \textit{monotonic}. A same conclusion can be drawn for the case of $s'_i-s_i > 1$  by mathematical induction.

For the latter, according to  the transition kernel defined in (\ref{probability}), one has
\begin{align}
 & \sum_{s^{+} \in \mathcal{S} } \mathcal{P}(s^{+}|s,a) \notag \\
 =  & \sum_{(s_i)^{+} \in \mathcal{S} } \mathcal{P}((s_i)^{+}|s_i,a) + \sum_{(s_i^{-})^{+} \in \mathcal{S} } \mathcal{P}((s_i^{-})^{+}|s_i^{-},a) \notag \\
 \leq & \sum_{(s_i)^{+} \in \mathcal{S} } \mathcal{P}((s_i)^{+}|s'_i,a) + \sum_{(s_i^{-})^{+} \in \mathcal{S} } \mathcal{P}((s_i^{-})^{+}|(s'_i)^{-},a) \notag \\
 = & \sum_{s^{+} \in \mathcal{S} } \mathcal{P}(s^{+}|s',a). \notag
\end{align}

Hence, the value function $V(\cdot)$ is \textit{monotonic} on $\mathcal{S}$.

\vspace{6pt}

\textbf{Part 2: (\textit{Submodular})} By utilizing same techniques, it can be shown that  $V(\cdot)$ is \textit{submodular} on $\mathcal{S}$.

Hence, the proof of Theorem \ref{thm2} is complete.

\bibliographystyle{plain}
\bibliography{ref}

\end{document}